\documentclass[12pt]{iopart}
\usepackage{iopams}
\usepackage[ruled]{algorithm}
\usepackage{algorithmic}
\usepackage{epsfig}
\usepackage{graphicx}
\usepackage{multirow}
\usepackage[sort&compress,numbers]{natbib}
\usepackage{amstext}

\newcommand{\eqref}[1]{(\ref{#1})}

\def\bA{\boldsymbol{A}}

\begin{document}

\title[On the entropy of protein families]{On the entropy of protein families}

\author{John P.~Barton$^{1\text{-}3,6}$, Arup K.~Chakraborty$^{1\text{-}6}$, Simona Cocco$^7$, Hugo Jacquin$^7$,  R\'emi Monasson$^8$}

\address{
$^1$ Ragon Institute of MGH, MIT \& Harvard, Cambridge, MA, USA \\
$^2$ Department of Chemical Engineering, Massachusetts Institute of Technology, Cambridge, MA, USA \\
$^3$ Department of Physics, Massachusetts Institute of Technology, Cambridge, MA, USA \\
$^4$ Department of Chemistry, Massachusetts Institute of Technology, Cambridge, MA, USA \\
$^5$ Department of Biological Engineering, Massachusetts Institute of Technology, Cambridge, MA, USA \\
$^6$ Institute for Medical Engineering \& Science, Massachusetts Institute of Technology, Cambridge, MA, USA \\
$^7$ Laboratoire de Physique Statistique de l'ENS, UMR 8550, associ\'e au CNRS et \`a l'Universit\'e P\&M. Curie, 24 rue Lhomond, 75005 Paris, France \\
$^8$ Laboratoire de Physique Th\'eorique de l'ENS, UMR 8549, associ\'e au CNRS et \`a l'Universit\'e P\&M. Curie, 24 rue Lhomond, 75005 Paris, France 
}
\eads{
\mailto{cocco@lps.ens.fr}
}

\begin{abstract}
Proteins are essential components of living systems, capable of performing a huge variety of tasks at the molecular level, such as recognition, signalling, copy, transport, ... The protein sequences realizing a given function may largely vary across organisms, giving rise to a protein family. Here, we estimate the entropy of those families based on different approaches, including Hidden Markov Models used for protein databases and inferred statistical models reproducing the low-order (1- and 2-point) statistics of multi-sequence alignments. We also compute the entropic cost, that is, the loss in entropy resulting from a constraint acting on the protein, such as the fixation of one particular amino-acid on a specific site, and relate this notion to the escape probability of the HIV virus. The case of lattice proteins, for which the entropy can be computed exactly, allows us to provide another illustration of the concept of cost, due to the competition of different folds. The relevance of the entropy in relation to directed evolution experiments is stressed. 
\end{abstract}


\maketitle

\section{Introduction}
Characterizing the statistical properties of a family of homologous protein sequences is a problem of fundamental importance in genomics. It is well known for instance that the frequencies of amino acids vary substantially along the sequence from site to site, as residues are generally strongly conserved in the protein cores and in binding pockets \cite{eddybook,consurf}. As the number of available sequences has hugely increased over the last years, higher-order statistical properties may now be accurately estimated. Correlations between pairs of residues in the sequence are known to reflect structural, functional, or phylogenetic constraints acting on the protein sequences \cite{lapedes99,rausell2010,panzos97,valentia2013}. Conservation, pairwise correlations, and possibly higher-order statistical constraints limit the number of putative proteins in a given family. It is of fundamental interest from an evolutionary point of view to be able to quantitatively estimate the diversity of proteins corresponding to a given family, and by extension, sharing the same biological function. The present paper, based on a variety of modeling approaches and of sequence data, is a modest attempt in this direction.

A natural way to quantify the diversity of proteins with the same function is through the Gibbs-Shannon entropy of the distribution of sequences in the corresponding protein family. Qualitatively, this entropy can be thought of as the logarithm of the number of sequences in the family, though there need not be a sharp divide between functional sequences (those belonging to the family) and dysfunctional ones. In the course of evolution, Nature has sampled many protein sequences across largely diverse organisms. Natural selection weeds out dysfunctional sequences, while amplifying those that perform their function efficiently. Current databases such as UniProt or PFAM \cite{pdb,uniprot,pfam} give us a sample of the diversity of those good sequences, {\em i.e.}~ones that ensure large fitnesses to the organisms compared to other protein sequences. However, despite massive sequencing efforts the number of available sequences is likely to be incredibly small compared to all possible sequences with high fitnesses. That is, we only observe a subset of the true distribution of functional sequences. We are thus faced with the difficult task of estimating the entropy of a probability distribution over the sequence space in the presence of dramatic undersampling. This is only possible under strong assumptions on the smoothness of the sequence distribution. Here we explore several different approaches for estimating the entropy for protein families, given a limited sampling of sequences. 

One popular approach in this context is to consider Maximum Entropy distributions \cite{Jaynes:1982wh,bialekbook,Weigt:2009ba,burger10,balakrishnan11} reproducing low-order statistics of the amino acids in the sequence databases, generally the single-site and pairwise frequencies. The corresponding distributions are smooth in the sense that they correspond to the Gibbs distributions associated to Potts Hamiltonians with local fields and pairwise couplings only. The difficulty in this approach is to compute those interaction parameters from the sequence statistics, and the corresponding entropies. In the present paper, we will resort to an approximate method allowing us to access those quantities, the Adaptive Cluster Expansion developed in \cite{Cocco:2011fo,Cocco:2012id}, and will apply it to real protein data (homologous protein families in Section \ref{sec:ace} and HIV sequence data in Section \ref{sec:hiv}) and to synthetic, lattice-based protein models (Section \ref{sec:LP}) \cite{gutin90}. This method estimates the cross-entropy between the inferred Potts model and the data, which is equal to the entropy of the Potts model that reproduces the desired statistics from the data. In addition to giving us access to absolute estimates of the entropies of the protein families, our approach allows us to compute changes of entropies related to additional constraints acting on the proteins. To illustrate this concept in the case of HIV, we will compute the variation in entropy as one amino acid is fixed to its consensus value. The loss in entropy, or entropy cost associated to this local constraint, is naturally related to the escape probability of a pathogen (virus or bacterium) from a drug or an immune attack. The latter can force mutations on one or multiple sites, and largely decreases the availability of putative escaping sequences. Another illustration will be provided by lattice-based proteins, where we will measure the decrease in entropy of a protein family, defined as the set of sequences folding properly into a given structure, resulting from the introduction of competing structures, {\em i.e.}~alternative folding conformations. 

We also estimate the entropy of the protein families in PFAM using their associated Hidden Markov Models (HMM) \cite{eddybook}. HMM define protein profiles, which are used to classify the families and answer sequence queries. HMM are, to some extent, similar to Maximum Entropy models reproducing 1-point statistics only, that is, to non-interacting Potts models with local fields only. However, HMM are capable of handling sequences of any length through the insertion of extra amino acids (for longer sequences than the length of the profile) or of gaps (for shorter sequences). As calculating exactly the value of the entropy of HMM models is generally a hard task, we will establish some bounds and approximations to this value in Section \ref{sec:HMM}. 

Last of all, in Section \ref{sec:discussion}, we summarize our findings, and compare the values of the entropies found with our different approaches, and to previous estimates in the literature \cite{Sha:1998}. We comment in particular the possible relevance of our results for directed evolution experiments, where protein sequences are evolved and selected in vitro, starting from a pool of random sequences.

\section{Wide-scale analysis of the entropy of HMM profiles across the PFAM database} \label{sec:HMM}

\subsection{Formalism for Hidden Markov Models}

Hidden Markov Models (HMM) are routinely used to define protein family profiles in databases, such as PFAM \cite{pfam}. The underlying principle for HMM is the existence of hidden states, which condition the set of symbols (amino acids or gaps) composing the sequence. Briefly speaking, an HMM jumps from one hidden state $\sigma$ to another state $\tau$ in a sequential and stochastic way, depending on a set of transition rates. After each transition to a new hidden state, a symbol $A$ may be produced, with an emission probability depending on the hidden state, and added to the sequence. When the last hidden state is reached the sequence is complete. A detailed description of HMM profiles can be found in \cite{eddybook}, Chapter 5. Hereafter we briefly expose their salient aspects and introduce some notations. 

In an HMM, for a profile of length $N$, the number of hidden states relevant to our purpose is $N_s=3N+1$. The initial and final states are denoted by, respectively, $B$ and $E$. In between $B$ and $E$ the model includes $N$ match states, denoted by $M_j$, with $j=1,2, \ldots , N$; $N-1$ insertion states $I_j$, with $j=1,2,\ldots , N-1$; $N$ deletion states $D_j$, with $j=1,2, \ldots, N$. Amino-acid symbols are emitted by the $I$ and $M$ states. Insertion states $I_j$ allow for the emission of excess symbols and to produce sequences longer than the profile length. Match states $M_j$ emit symbols with probabilities dependent on the position ($j$), and reflect the pattern of amino acid conservation along the profile. Deletion states $D_j$ represent a gap in the sequence, {\em i.e.}~the lack of correspondence to the site $j$ of the profile. Note that the number of insertion states, $N-1$, is different from the one ($=N+1$) in the description of HMMs in \cite{eddybook}, Chapter 5, Fig.~5.2; the reason is that our definition of HMMs corresponds to the one of the Matlab Bioinformatics toolbox we use to download the profiles from the PFAM database, and does not consider insertion states associated to the $B$ and $E$ states. 

An HMM is fully defined by the transition rate matrix ${\cal T}$, of size $N_s\times N_s$, which gives the probability of jumping from any hidden state $\sigma$ to another $\tau$, and by the emission matrix ${\cal E}$, which gives the probability of emitting symbol $A$ given the hidden state $\sigma$. If the hidden state is an insertion or match state, $\sigma=I_j$ or $M_j$, any of the 20 amino acids $A$ may be emitted with probability ${\cal E}(A|\sigma)$; if the hidden state is a deletion state, $\sigma=D_j$, the gap symbol is emitted with probability unity.
The weight of the sequence ${\bf A}=(A_1,A_2, \ldots, A_L)$ with $L$ emitted symbols is given by
\begin{equation}\label{pofx}
P({\bf A};L) = \sum_{\boldsymbol\sigma = (\sigma_1,\sigma_2, \ldots , \sigma_L)} {\cal T} (B\to \sigma_1) \prod_{\ell=1}^{L} \bigg[ {\cal T}(\sigma_\ell\to\sigma_{\ell+1}) {\cal E}(A_\ell | \sigma_\ell)\bigg]  \ ,
\end{equation}
where we have defined $\sigma_{L+1}\equiv E$. The sum over all sequences $\bf A$ of $P({\bf A};L)$ is the probability  to reach $E$ from $B$ through a path of length $L$ across the hidden states; this sum, denoted by $P(L)$, is {\em a priori} smaller than unity, {\em e.g.} if the path length $L$ is smaller than the model size $N$ and $E$ can not be reached from $B$. In practice, however, $P(L)$ converges to unity as soon as $L$ exceeds $N$, see below. Our goal is then to compute the entropy of the HMM,
\begin{equation}
S_1 (L)= -\sum _{\bf A} P({\bf A};L) \; \log P({\bf A};L)\ ,
\end{equation}
which is a function of the matrices ${\cal T},{\cal E}$ only. An exact calculation of $S_1(L)$ is very difficult (see below), and we will instead compute the lower bound
to the entropy,
\begin{equation}
S_1 (L)>  S_2(L)\ ,
\end{equation}
and the approximation
\begin{equation}
S_1 (L)\simeq 2\, S_2(L)- S_3(L)\ ,
\end{equation}
based on the Renyi entropies $S_q(L)$:
\begin{equation}
S_q(L) = \frac 1{1-q} \log\bigg[ \sum _{\bf A} P({\bf A};L)^q \bigg] \ .
\end{equation}
Note that $2\, S_2(L)- S_3(L)$ is not guaranteed to be a lower bound to $S_1(L)$, but is generally a closer estimate of $S_1(L)$ than the guaranteed lower bound $S_2(L)$.

We now turn to the computation of $S_q(L)$, where $q$ is integer valued. According to (\ref{pofx}), we have
\begin{eqnarray}\label{pofx2}
\sum_{\bf A} P({\bf A};L)^q &=& \sum_{\boldsymbol\sigma ^{(1)}  , \boldsymbol\sigma ^{(2)}, \ldots , \boldsymbol\sigma ^{(q)}} \prod_{m=1}^q\;{\cal T} (B\to \sigma^{(m)}_1) \prod_{\ell=1}^{L} \bigg[ \prod_{m=1}^q {\cal T}(\sigma ^{(m)}_\ell\to\sigma^{(m)}_{\ell+1})  \nonumber \\
&\times& R( \sigma ^{(1)} _\ell , \sigma ^{(2)}_\ell , \ldots , \sigma ^{(q)}_\ell) \bigg]  \ ,
\end{eqnarray}
where we have defined $\sigma _{L+1}^{(1)} =\sigma _{L+1}^{(2)}=...=\sigma _{L+1}^{(q)}=E$, and
\begin{equation}\label{pofx2b}
R( \sigma ^{(1)}  , \sigma ^{(2)}, \ldots , \sigma ^{(q)})  = \sum_{A}  {\cal E}(A | \sigma^{(1)})\; {\cal E}(A| \sigma^{(2)})\;  \ldots \;  {\cal E}(A| \sigma^{(q)})\ ,
\end{equation}
for any set of $q$ states $\sigma ^{(1)} , \sigma ^{(2)}, \ldots , \sigma ^{(q)}$,.
We introduce the indices $\hat{\sigma}=(\sigma ^{(1)}, \sigma ^{(2)},\ldots , \sigma ^{(q)})$ to label the elements of the  $(N_s)^q\times(N_s)^q$-dimensional effective transition rate matrix ${\cal M}_q$:  
\begin{equation}
{\cal M}_q \big(\hat{\sigma} \to \hat{\tau} \big) =\displaystyle{\prod_{m=1}^q{\cal T} ( \sigma^{(m)} \to \tau^{(m)})}\times  \left\{ \begin{array}{c}
 1 \ \hbox{\rm if} \ \hat{\sigma} = \hat B \ \hbox{\rm or}\ \hat{\tau} = \hat E\ ,\\
R( \sigma^{(1)}  , \sigma ^{(2)}\ldots , \sigma ^{(q)}) \ \hbox{\rm otherwise} \ ,  \\
\end{array}
\right. 
\end{equation}
where $\hat{B} = (B,B,\ldots , B)$, $\hat{E} = (E,E,\ldots , E)$.
Using those notations, we write
\begin{equation}\label{ppp99}
\sum_{\bf A} P({\bf A};L)^q  = {\cal M}_q^L\big( \hat B \to \hat E \big)\quad \hbox{\rm and}\quad S_q(L) = \frac 1{1-q} \log {\cal M}_q^L\big( \hat B \to \hat E \big) \ .
\end{equation}
where ${\cal M}_q^L$ denotes the $L^{th}$-matrix power of ${\cal M}_q$. This formula explains why the calculation of $S_q(L)$, with $q\ge 2$ is easier than the calculation of $S_1(L)$.
Indeed, the hard computational step in the calculation of the entropy is obviously the summation over the enormous set of sequences $\bf A$, whose size grows exponentially with the length $L$. For integer values of $q\ge 2$, the summation can be split in $L$ independent sums over the symbols $A_\ell$, which define the effective matrix $R$ in (\ref{pofx2b}). The Renyi entropies $S_q$ (with $q\ge 2$) can then be computed in a time growing linearly with $L$ (and not exponentially) from the knowledge of the $L^{th}$ power of the transition matrix ${\cal M}_q$ in (\ref{ppp99}). Unfortunately the size of ${\cal M}_q$ grows exponentially with $q$, and this trick is limited to small values of $q$. 

The formulas above were implemented in a Matlab routine. The rate  matrix ${\cal T}$ and the emission matrix ${\cal E}$ were downloaded from PFAM profiles, and used to compute the ${\cal M}_2$ and ${\cal M}_3$ matrices. As the size of the latter grows as the cubic power of the number $N_s$ of hidden states the computation of the Renyi entropy $S_3(L)$ was done for moderate profile length only, {\em i.e.}~in practice $N\le 100$.

\subsection{Convergence with $L$}

\begin{figure}
\begin{center}
\includegraphics[width=7cm]{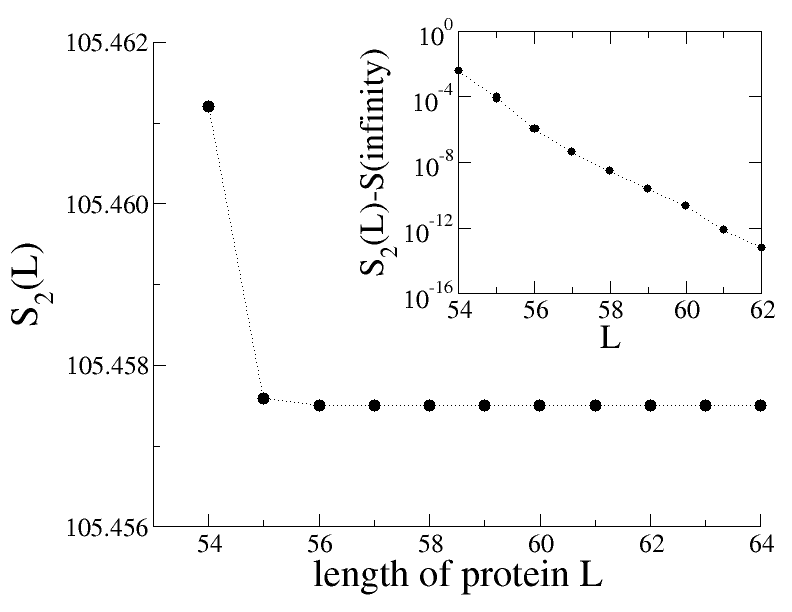}
\includegraphics[width=7cm]{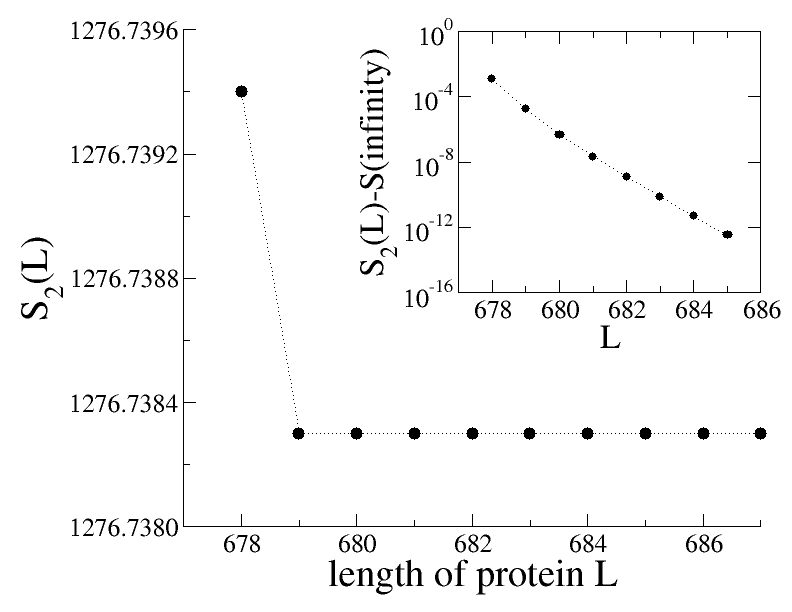}
\caption{Lower bounds $S_2(L)$ to the entropy of the Hidden Markov Models for families PF00014 ({\bf A}) and PF00063 ({\bf B}) as functions of the length of the proteins emitted by the models. For both families, the entropy has a finite value when $L$ exceeds the profile length (53 for PF00014 and 677 for PF00063). Insets: difference between $S_2(L)$ and its asymptotic value vs. $L$.
}
\label{fig:conv}
\end{center}
\end{figure}

We first study how our low bound for the HMM entropy depends on the length $L$ of `emitted' proteins. We plot in Fig.~\ref{fig:conv} the value of the Renyi entropy $S_2(L)$ as function of $L$ for two families: PF00014, a trypsin inhibitor, also studied in Section~\ref{sec:trypsin} and PF00063, a myosin-head protein. 
Those two families were chosen for the very different values of the lengths of their profiles: $N=53$ for PF00014 and $N=677$ for PF00063. The entropy $S_2(L)$ is equal to minus infinity as long as $L\le N$. The reason is that, in PFAM HMMs, the probabilities of transitions from any Match state $M_j$ (with $j< N$) to the end state $E$ are zero. $E$ can therefore not be reached in less than $L=N+1$ steps. 
The value of the entropy $S_2(L=N+1)$ corresponds to the shortest transition path (through the Match or Deletion states) connecting $B$ to $E$. As $L$ increases, the probabilities of more and more processes (including self-transitions of Insertion states onto themselves, which have low but non-zero probabilities \cite{eddybook}) are collected, and the sum of the squared probabilities increases, which makes $S_2$ decrease. 
Note that, once the state $E$ is reached the system enters an infinite loop (through the transition $E\to E$) and $S_2(L)$ does include all the contributions coming from paths connecting $B$ to $E$ with length shorter or equal to $L$ (by convention, in the calculation of $R$ in (\ref{pofx2b}), we consider that $E$ emits empty symbols). We see that the entropy reaches an asymptotic plateau very quickly as $L$ increases above $N+1$ (Inset of Fig.~\ref{fig:conv}). 
In practice we choose $L=1.2\times N$ to be sure that the convergence has been reached, and all paths from $B$ to $E$ have been taken into account ($P(L)=1$). A similar behaviour is observed for the Renyi entropy of order 3 as a function of the protein length $L$ (not shown). To lighten notations we write in the following $S_q$ for our estimate of the asymptotic entropy $S_q(L\to\infty)$.

\subsection{Fixed-Length Model built from HMM}\label{flames}

\begin{figure}
\begin{center}
\includegraphics[width=10cm]{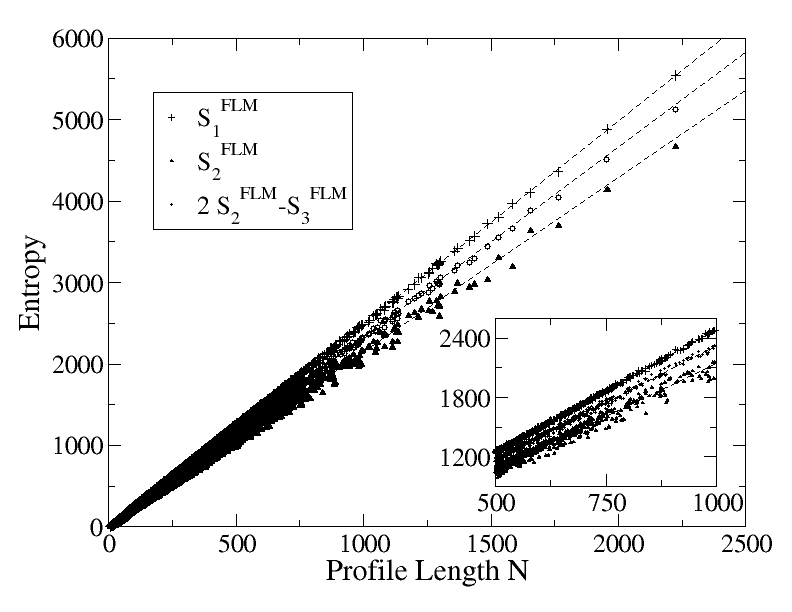}
\caption{Entropies of the Fixed-Length Model across the 16,229 families in PFAM~28.0 vs. the length $N$ of the profiles of the families. Pluses: $S_1^{FLM}$; Circles: approximation $2S_2^{FLM}-S_3^{FLM}$; Triangles: lower bound $S_2^{FLM}$. The continuous lines show the linear fits (\ref{linearflm}), (\ref{linearflm2}) and (\ref{linearflm3a}).
 Inset: magnification of the plot in the region $500\le N\le 1000$.
}
\label{fig:flm}
\end{center}
\end{figure}

As a first step we ignore the possibility of insertion and deletion. The resulting simplied HMM model consists of the $N$ Match states, visited one after the other in a sequential way. 
On each Match state $M_j$, an amino acid $A_j$ is emitted according to the local probability of emission ${\cal E}$. In this simple Fixed-Length Model (FLM), symbols are emitted independently of each other. The entropy of the distribution of sequences produced with the FLM is therefore
\begin{equation}
S_{1}^{FLM} = - \sum _{j=1}^N \sum_{A_j} {\cal E} (A_j|M_j) \log {\cal E}(A_j|M_j) \ .
\end{equation}
In Fig.~\ref{fig:flm} we show the entropy $S_1^{FLM}$ of the FLM for the 16,229 families PF$nnnnn$ in the PFAM 28.0 database, released in May 2015, with numbers $nnnnn$ smaller or equal to 17,126. A linear fit of the entropy as a function of the length $N$ of the profile is excellent and gives,
\begin{equation}\label{linearflm}
S_1^{FLM} \simeq \sigma_1 ^{FLM} \times N \ , \quad \hbox{\rm where} \quad \sigma_1 ^{FLM} = 2.4880 \pm 0.0006  \ ,
\end{equation}
with $95\%$ confidence.

\begin{figure}
\begin{center}
\includegraphics[width=10cm]{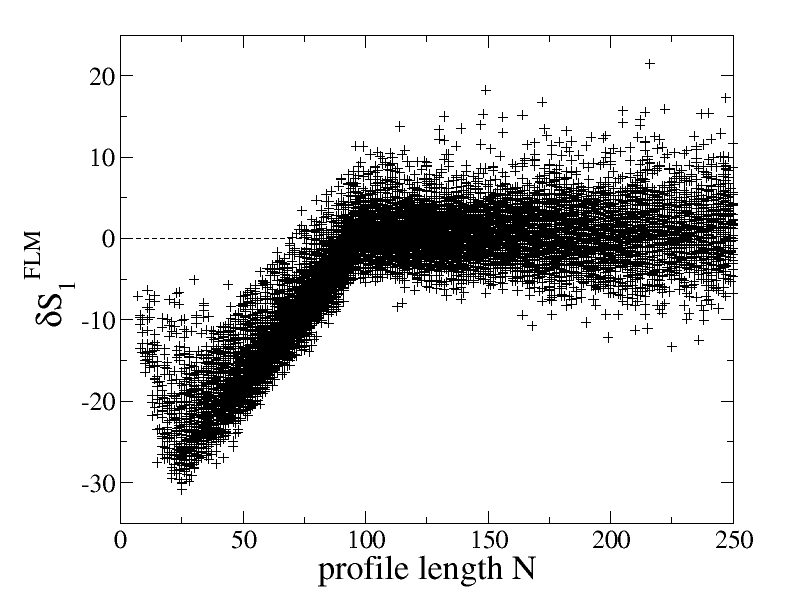}
\caption{Difference between the entropy $S_1^{FLM}$ of the Fixed-Length Model and its linear approximation (\ref{linearflm}), across the 16,229 families in PFAM~28.0, vs. the length $N$ of the profiles of the families. }
\label{fig:devflm}
\end{center}
\end{figure}

To investigate the finer statistics of the entropy as a function of $N$, we plot in Fig.~\ref{fig:devflm} the residuals of the linear fit,
\begin{equation}\label{linearflmb}
\delta S_1^{FLM} = S_1^{FLM} - \sigma_1 ^{FLM} \times N  \ .
\end{equation}
We observe a systematic and negative deviation for lengths $N<100$.  The reason for this variation is not entirely clear; one possible explanation could be the extra penalty introduced for shorter profiles \cite{finn2009}. This penalty term is intended to remove the sequences, whose scores barely exceed the level of noise expected for random sequences. As a result, the shorter the profile length, the more sequences are removed, with a net effect of reducing the entropy of the family.
For larger sizes, the deviation is on average zero, with a standard deviation comprised in a strip of width growing as the square root of $L$. 
This result is expected for independent-site models, due to the central limit theorem.

We also plot in Fig.~\ref{fig:flm} the lower bound $S_2^{FLM}$ and the approximation $2S_2^{FLM}-S_3^{FLM}$ to the entropy $S_1^{FLM}$. As the value of $S_1^{FLM}$ is exactly known, we can assess the accuracy of the bound and of the approximation.
This will be useful below in the case of HMM, where an exact computation of $S_1$ is out of reach. We observe that both quantities increase on average linearly with the profile length $N$, with the slopes
\begin{equation}\label{linearflm2}
S_2^{FLM} \simeq \sigma_2 ^{FLM} \times N \ , \quad \hbox{\rm where} \quad \sigma_2 ^{FLM} = 2.1438 \pm 0.0012 \ ,
\end{equation}
and
\begin{equation}\label{linearflm3a}
2\;S_{2}^{FLM} -S_3^{FLM} \simeq \sigma_{2-3} ^{FLM} \times N \ , \quad \hbox{\rm where} \quad \sigma_{2-3} ^{FLM} = 2.3265 \pm 0.0006  \ ,
\end{equation}
both with $95\%$ confidence. The deviations of the Renyi entropies $S_2^{FLM}$ and $S_3^{FLM}$ with respect to those linear fits show roughly the same behaviour as in the $S_1^{FLM}$ case (not shown). However, for the latter entropies, deviations are larger and not Gaussianly distributed, as the central limit theorem does not apply to Renyi entropies of order different from unity.

\subsection{Bounds and approximation for the full HMM model}

\begin{figure}
\begin{center}
\includegraphics[width=14cm]{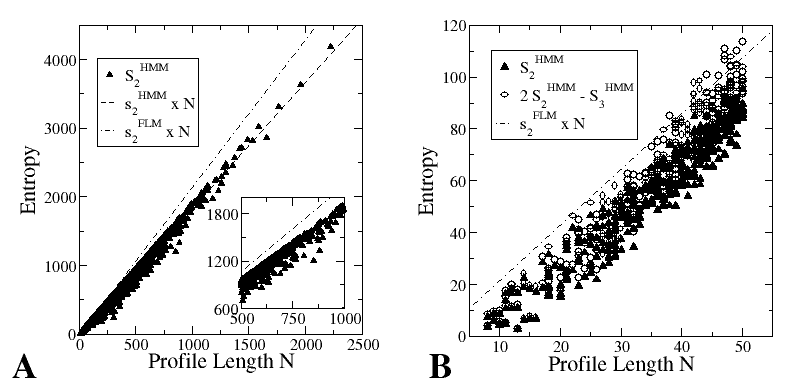}
\caption{Entropies of the Hidden Markov Model across the 16,229 families in PFAM database vs. the length $N$ of the profiles of the families. {\bf A.} Lower bound $S_2^{HMM}$ (triangles). The lines show the linear fit (\ref{linearhmm2}) and the one for the Fixed-Length Model, see (\ref{linearflm2}).  Inset: magnification of the plot in the region $500\le N\le 1000$. {\bf B.} Profiles with lengths $N\le 50$ only across all families of index $< 5000$ in PFAM 24.0. Circles: approximation $2S_2^{HMM}-S_3^{HMM}$; Triangles: lower bound $S_2^{HMM}$.
}
\label{fig:hmm}
\end{center}
\end{figure}

We plot in Fig.~\ref{fig:hmm}A the lower bound $S_2^{HMM}$ to the true entropy $S_1^{HMM}$ of the HMM model, which we are not able to compute exactly. We observe that $S_2^{HMM}$ increases on average linearly with the profile length $N$, with the slopes
\begin{equation}\label{linearhmm2}
S_2^{HMM} \simeq \sigma_2 ^{HMM} \times N \ , \quad \hbox{\rm where} \quad \sigma_2 ^{HMM} = 1.8367 \pm 0.0015  \ ,
\end{equation}
within 95\% accuracy. The slope is 14.3\% lower than its counterpart in the Fixed-Length Model. However, a refined approximation of the entropy based on the calculation of the Renyi entropy of order 3 gives
\begin{equation}\label{linearhmm3}
2\;S_{2}^{HMM} -S_3^{HMM} \simeq \sigma_{2-3} ^{HMM} \times N \ , \quad \hbox{\rm where} \quad \sigma_{2-3} ^{FLM} = 2.236 \pm 0.008  \ ,
\end{equation}
which is only 4\% less than the slope found for the Fixed-Length Model, see plot in Fig.~\ref{fig:hmm}B. Those results suggest that the entropy of the HMM is only weakly affected by the non-independence between the symbols due to the presence of gaps, and is very close to its FLM counterpart.
 
\subsection{Comparison of HMM entropies for two distributions of PFAM}

\begin{figure}
\begin{center}
\includegraphics[width=8cm]{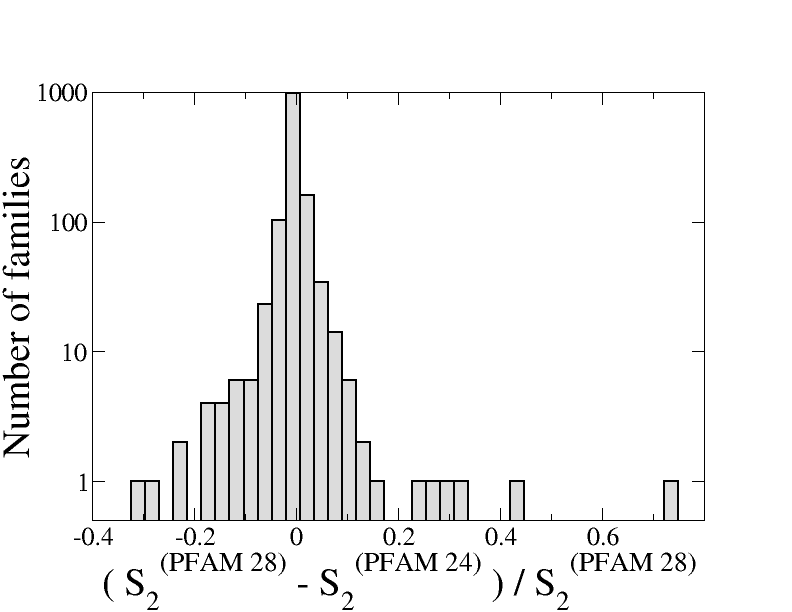}
\caption{Histogram of relative changes in the Renyi entropies $S_2$ between PFAM releases 24.0 and 28.0 for more than 1,300 protein families with unchanged profile lengths.
}
\label{fig:changeS2}
\end{center}
\end{figure}

Hereafter we study how the changes in the HMM from one PFAM release to another affect the value of the entropy. To do so we consider the current PFAM 28.0 release (May 2015) and release 24.0 (October 2009).  To be as conservative as possible in our estimate of the change in entropy, we first identify 1343 families (among the familiies PF$nnnnn$, with $nnnnn<5000$ in release 24.0), whose profile length have not changed from one release to another. 
The histogram of the relative variations of Renyi entropy $S_2$ is shown in Fig.~\ref{fig:changeS2}. The histogram is centered in zero, and is roughly symmetric around the origin. About 2\% of the families, that is, 28 families, show a relative change in entropy larger than 10\%. Those large changes show that some families are still affected by strong undersampling. Once rescaled by the length $N$ of the profiles, the variations in entropy range between -0.18 and 0.08, which represent variations of about 5 to 10\% of the average slope $\sigma_2^{HMM}$  computed above.

\section{Potts models for protein families and inference with the Adaptive Cluster Expansion algorithm} \label{sec:ace}

\subsection{Cross-entropy and general formalism}

In this Section we use a Potts model to fit the probability distribution of the sequences $\bA=(a_1,a_2,\ldots,a_N)$ associated to a specific protein family. The Potts distribution naturally arises in the Maximum Entropy framework as the least constrained (maximum entropy) distribution capable of reproducing the set $\bf p$ of single-site and pairwise frequencies of amino-acids in the natural multi-sequence alignment (MSA) of a given protein family. The parameters of the Potts model are the local fields ${\bf h}=\{h_{i}(a)\}$, which may be interpreted as position weight matrices, and the couplings ${\bf J}=\{J_{ij}(a, b)\}$ between the amino acids $a$ and $b$ at the sites $i$ and $j$. Those parameters have to be fitted to reproduce the pairwise frequencies $p_{ij}(a,b)$ and the single-site frequencies $p_i(a)$ computed from the MSA.

The Potts parameters can be inferred through the minimization of the cross-entropy between the model (defined by its parameters ${\bf J}, {\bf h}$) and the data ($\bf p$), equal to minus the log-likelihood of the sequence data:
\begin{eqnarray} 
\label{eq:Pottsentropy}
S^{cross}({\bf J},{\bf h}| {\bf p }) &=& \log\left[ \sum_{\bA} \exp\left(\sum_i h_i(a_i) + \sum_{i<j} J_{ij}(a_i, a_j)\right) \right] \nonumber \\
&-& \sum_{i}\sum_{a=1}^{q_i} h_i(a) p_i(a) -\sum_{i<j}\sum_{a=1}^{q_i}\sum_{b=1}^{q_j} J_{ij}(a,b) p_{ij}(a,b) .
\end{eqnarray}
In the expression above, $q_i$ is the number of Potts symbols on site $i$. The maximum value of $q_i$ is 21 (20 amino acids plus the gap symbol), but $q_i$ can be sizeably smaller on sites where only a few amino acids appear in the MSA. The cross-entropy $S^{cross}$ is equivalent to the entropy $S^{Potts}$ of the Potts model reproducing the 1- and 2-point statistics of the MSA (modulo two contributions introduced below). Hence, the cross-entropy can be used to quantify the diversity of sequences in the protein family in the same way as the entropy of the HMM in Section~\ref{sec:HMM}. To make sure that the minimum $\bf J,h$ is finite and unique we add to $S_{cross}$ above the $L_2$-norm regularization term
\begin{equation}\label{dl2}
\Delta S ^{L_2}({\bf J},{\bf h})= \frac {1} {200\gamma } \sum_{i,a} h_i(a) ^2 + \frac 1 {2\gamma}\sum_{i<j} \sum_{a,b} J_{ij}(a,b) ^2 \ ,
\end{equation} 
where $\sqrt \gamma$ is the magnitude of the largest couplings tolerated in the inference; the fields are allowed to take ten times larger values.
The Adaptive Cluster Expansion (ACE) introduced in \cite{Cocco:2011fo,Cocco:2012id,Barton:2014ek} is a way to calculate the minimal value of $S_{cross}+\Delta S^{L_2}$ over $\bf J$ and $\bf h$ through the construction and the summation of many clusters of strongly interacting sites.

We will investigate how the approximate value of the entropy calculated with the ACE procedure depends on the detailed procedure followed to format the data. In particular, this formatting includes
\begin{itemize}
\item the reweighting \cite{Morcos:2011c} of similar sequences in the MSA to reduce the phylogenetic correlations of the sequences. The weight of each sequence is taken to be the inverse of the number of sequences with a Hamming distance smaller than $wN$ (this number is always larger or equal to one, as the sequence itself is included). Hereafter, we will compare results obtained for $w=0.2$ and $w=0$ (no reweighting);
\item the regularization term (\ref{dl2}) with $\gamma\simeq {1\over M}$, where $M$ is the number of sequences in the MSA, which can be tuned to improve the convergence of the ACE;
\item the reduction of the Potts alphabet. To speed up the algorithm and avoid overfitting we reduce the number of Potts symbols $q_i$ on each site. To do so we consider only the observed amino acids as possible Potts symbols, and we may also lump together those which are not frequently observed in a unique, abstract Potts symbol. More precisely, 
we use a criterion based on the frequency, and group together all the amino acids observed with probability $p<p_{red}$ or whose contributions to the site entropy is smaller than a fraction $S_{red}$. We will compare the efficiencies of both criteria for different values of the reduction parameters $p_{red}$ and $S_{red}$. 
\item the effect of the substitution of the gaps in the MSA with amino acids, randomly drawn according to their frequencies in the sequences without gaps at each position. 
\end{itemize} 

\subsection{Application to families PF00014 and PF00397} \label{sec:trypsin}
 
The inference procedure above will be applied to two small proteins: the trypsin inhibitor (PFAM family PF00014) and WW (PF00397). Trypsin inhibitor is a protein domain reducing the activity of trypsin, an enzyme involved in the breakdown of proteins during the digestion; its PFAM profile includes 53 sites. It has been used as a benchmark for structural prediction based on amino-acid covariation \cite{Morcos:2011c,aurell2013,cocco2013}. WW is a protein domain able to bind peptides, and composed of 31 amino acids. WW was used as a benchmark to test the success of covariation-based procedures to design new folding and biologically functional sequences \cite{russ05,Socolich:2005js}. We will compare the results of the Potts inference obtained from the MSA in PFAM releases 24.0 and 27.0.
 
\begin{figure}
\begin{center}
\includegraphics[width=10cm]{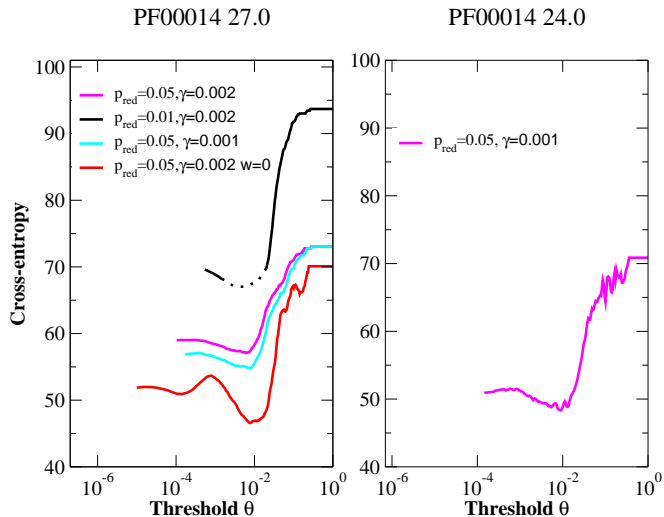}
\caption{PF00014: Cross-entropy obtained by the ACE algorithm as a function of the threshold $\theta$ for selecting clusters, for the MSA of PFAM 27.0 (left) and 24.0 (right). 
Amino acids with $p<p_{sel}$ are regrouped in a single Potts state, and the regularization strength $\gamma$ is varied.
The reweighting factor is $w=0.2$ for all curves but one, where $w=0$, see legend. } 
\label{fig:entroPF00014pred}
\end{center}
\end{figure}

\begin{figure}
\begin{center}
\includegraphics[width=10cm]{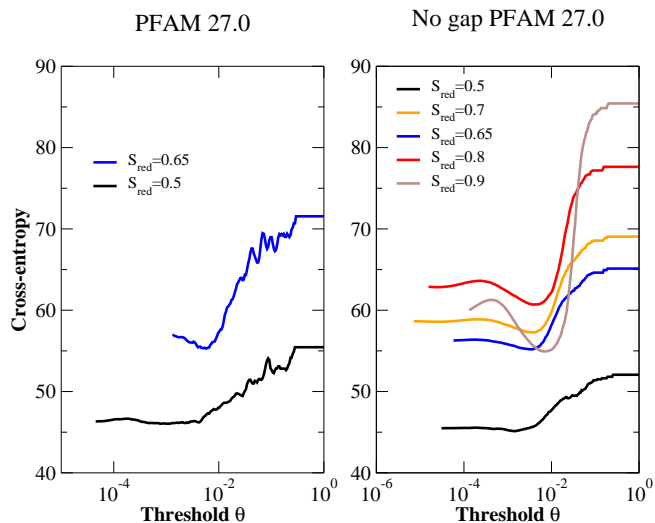}
\end{center}
\caption{PF00014: Cross-entropy obtained by the ACE algorithm as a function of the threshold $\theta$ for selecting clusters, for the MSA of PFAM 27.0 (left) and after removal of gaps through the randomization procedure (right). The reweighting factor is $w=0.2$. Potts state reduction according to the entropy-based criterion, with cut-off $S_{red}$. }
\label{fig:entroPF00014Sred}
\end{figure}

\begin{figure}
\begin{center}
\includegraphics[width=10cm]{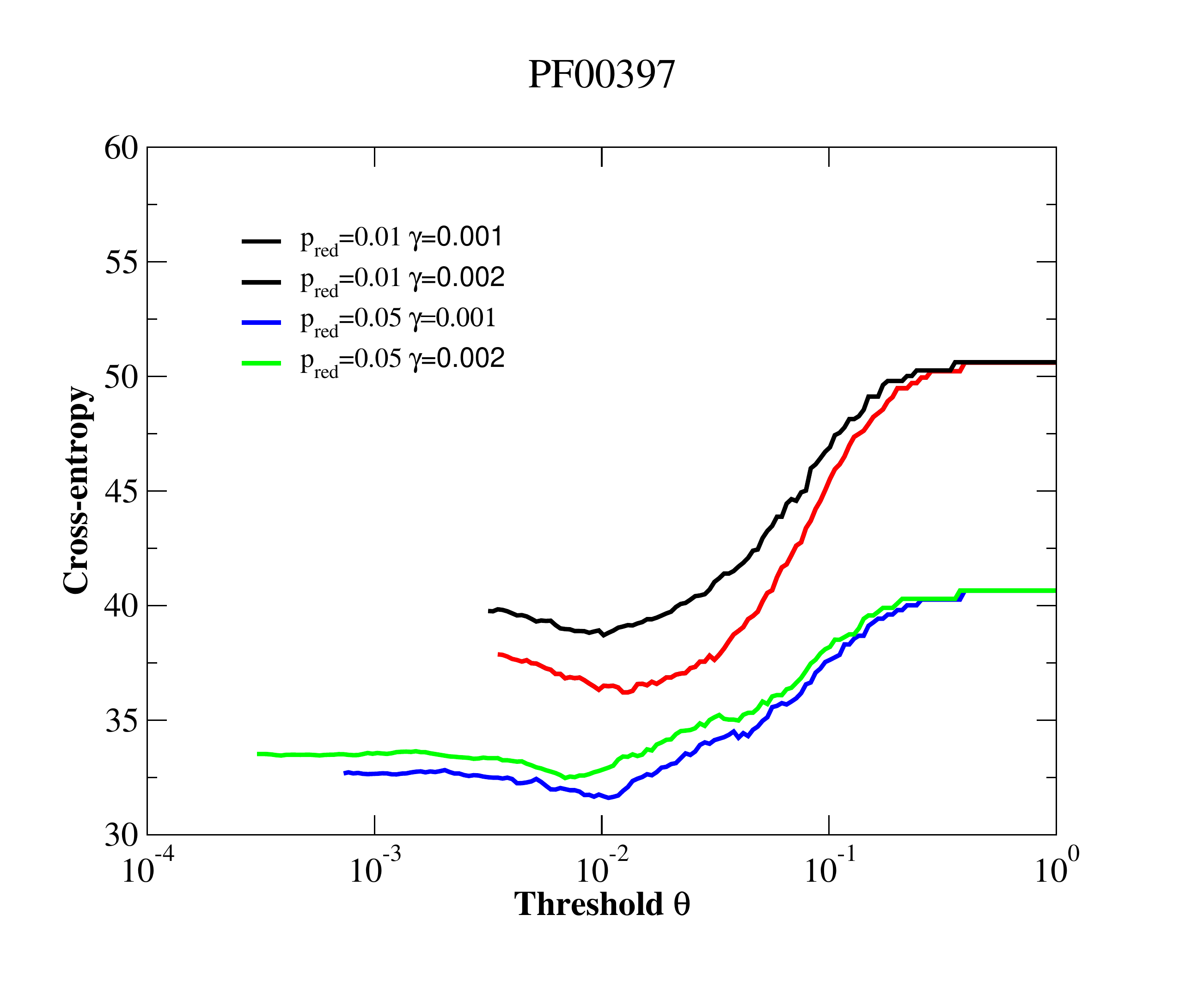}
\end{center}
\caption{PF00397: Cross-entropy obtained by the ACE algorithm as a function of the threshold $\theta$ for selecting clusters, for the MSA of PFAM 27.0, with reweighting $w=0.2$.  
Amino acids with $p<p_{sel}$ are regrouped in a single Potts state, and the regularization strength $\gamma$ is varied.
The reweighting factor is $w=0.2$.}
\label{fig:entroPF00397pred}
\end{figure}

In Figs.~\ref{fig:entroPF00014pred}, \ref{fig:entroPF00014Sred} and \ref{fig:entroPF00397pred} we show the entropies for the two proteins as functions of the threshold $\theta$ used to select clusters in the ACE procedure, of the number of Potts symbols with the frequency or the entropy criteria, and of the regularization strength $\gamma$. At the starting threshold at $\theta=1$ only single-site clusters are selected, which corresponds to an independent-site model (IM), and the cross-entropy reads
\begin{equation}\label{defsim45}
S^{IM}=-\sum_{i} \sum_{a=1}^{q_i} p_{i}(a) \log {p_{i}(a) } \ .
\end{equation}
Upon lowering the selection threshold $\theta$ more and larger clusters are summed in the expansion. The entropy decreases, possibly with some oscillations, and eventually converges at small threshold. We note that such oscillations are possible because the algorithm produces an estimate for the entropy by summing up the contributions from many small clusters of sites, which can be either positive or negative (for details see \cite{Cocco:2011fo,Cocco:2012id}). Once the entropy has reached convergence, we substract from its value the contribution $\Delta S^{L_2}$ coming form the regularization, see (\ref{dl2}), 
and add a contribution to partially correct for the clustering of amino acids in a unique Potts state,
\begin{equation}
\Delta S ^{AGG}= - \sum_{i} \sum_{a=1}^{k_i} \; p_i(a) \, \log \left( \frac{p_{i}(a)} {p_{i} (r)} \right ) \ .
\end{equation} 
In the expression above, $p_i(r)$ denotes the frequency of the abstract Potts symbol, $r$, which stands for the $k_i$ Potts states lumped together. By definition, $p_i(r)=\sum_{a=1}^{k_i} p_i(a)$. The correction $\Delta S ^{AGG}$ vanishes if no amino acid have been grouped on site $i$, and $p_i(r)=0$. If $k_i\ge 1$, $\Delta S ^{AGG}$ is not equal to zero, and is equal to the entropy of $k_i$ independent symbols with probabilities $p_{i}(a) /p_{i} (r)$, weighted by the probability $p_i(r)$ of the abstract Potts symbol. It allows us to recover the full model from the reduced one in the IM case, see Table~\ref{table:spf00014ind}. The final expression for the entropy is therefore $S^{Potts}=S^{cross}-\Delta S^{L_2}+\Delta S ^{AGG}$.

As a general trend, we observe that the cross-entropy decreases when the reduction parameter $p_{red}$ is made smaller or the cut-off fraction $S_{red}$ is made larger, see Tables \ref{table:spf00014pred}, \ref{table:spf00014ind}, \ref{table:spf00014Sred}, \ref{table:spf00397ind} and \ref{table:spf00397pred}. In other words, as the number of retained Potts symbols increases, the entropy decreases. This behaviour is easy to understand: keeping more Potts symbols amounts to reproducing more single-site and pairwise frequencies, and hence, to fulfil more constraints. Note also that due to the removal of the regularization contribution to the entropy, $\Delta S^{L_2}$, the value of the cross-entropy depends only weakly on the regularization strength $\gamma$, see results for $\gamma=0.001$ and $\gamma=0.002$ in Figs.~\ref{fig:entroPF00014pred} and~\ref{fig:entroPF00397pred}. Nevertheless, we observe that the cross-entropy increases slightly with $\gamma$, as the correlations are effectively less tightly constrained.
 
For PF00014, we obtain the entropy of the Independent Model, $S^{IM}=96.75$, and a corresponding entropy per site $\sigma^{IM}=1.79$ (Table~\ref{table:spf00014ind}). Naturally, as pairwise constraints are introduced the entropy of the corresponding model decreases. The entropy of the Potts model including couplings is $S^{Potts}= 67.68$ ($\sigma=1.28$) for $p_{red}=0.01$, and $S^{Potts}= 75.4$ ($\sigma=1.42$) for $p_{red}=0.05$ (Table~\ref{table:spf00014pred}). The decrease in entropy is similar when regrouping the symbol according to the $p_{red}$ or $S_{red}$-based criteria as long as the number of Potts symbols remaining are equivalent (Table~\ref{table:spf00014Sred}). For instance, results for the reduction $p_{red}=0.05$ are similar to the ones with $S_{red}=0.7$, and the ones with $p_{red}=0.01$ are similar to the ones with $S_{red}=0.8-0.9$. As noted above, enforcing more pairwise constraints (i.e.~taking smaller $p_{red}$ or larger $S_{red}$) results in lower entropy.
 
We have also calculated the entropy in the family PF00014 from PFAM 24.0, containing $M\simeq 2000$ sequences and an effective number of sequences $M_{eff}=1000$ after reweighting with $w=0.2$, and compared it to the outcome of the calculation for PFAM 27.0, corresponding to $M \simeq 4000$, $M_{eff}=2000$. We find that the entropy has increased by about 5\% between the two releases, probably a result of the introduction of more diverse sequences in the database. In the absence of reweighting ($w=0$), the entropy decreases (by about 4\%), as similar sequences are given more weights and the distribution of sequences is more peaked, see Table~\ref{table:spf00014pred}. For PF00397, we obtain a similar behavior: $S^{IM}=52.38$ $(\sigma=1.69)$ for the Independent Model, and $S^{Potts}= 38$ ($\sigma=1.22$) for $p_{red}=0.01$, $S^{Potts}= 43.82$ ($\sigma=1.41$) for $p_{red}=0.05$, see Tables \ref{table:spf00397ind} and \ref{table:spf00397pred}.

\begin{table} \begin{center}
\begin{tabular}{|c|c|c|c|c|}
\hline 
$p_{red},\gamma,w $ & Entropy $S^{Potts}$ & Cross-entropy $S^{cross}$ & $\Delta S^{L_2}$ & $\Delta S^{AGG}$  \\
\hline\hline	PFAM 24.0						& 		& 		&		&		\\
\hline 	0.01								& 67.68	& 69.61	& -5.00	& 3.07	\\
\hline 	0.05								& 81.02 	& 59.00	& -1.66	& 23.7 	\\
\hline	0.05, $\gamma=10^{-3}$			& 79.11	& 56.89	& -1.48	& 23.7 	\\
\hline 	0.05, $\gamma=10^{-3}$, $w=0$	& 75.41	& 51.88	& -2.74	& 26.28	\\
\hline\hline	PFAM 27.0						& 		& 		& 		&  		\\
\hline	0.05 							& 74.15	& 50.89	& -3.79	& 27.05	\\
\hline
\end{tabular}
\caption{PF00014. Results of the Potts model inference with the ACE procedure. Potts states were grouped together according to their frequencies, with cut-off $p_{red}$. Unless otherwise specified the reweighting factor is $w=0.2$ and the regularization strength $\gamma=0.002$. }
\label{table:spf00014pred} \end{center}
\end{table} 

\begin{table}
\begin{center}
\begin{tabular}{|c|c|c|c|c|}
\hline $p_{red}$ & Entropy $S^{IM}$ & Cross-entropy $S^{IM-cross}$ & $\Delta S^{L_2}$ & $\Delta S^{AGG}$  \\
\hline  0.01& 96.75&93.68&-0.014& 3.07 \\
\hline  0.05 & 96.75 &73.06&-0.01 &23.7 \\
\hline
\end{tabular}
\end{center}
\caption{PF00014, release 27.0. Independent model with selection of the number of Potts states. }
\label{table:spf00014ind}
\end{table} 
 
\begin{table}
\begin{center}
\begin{tabular}{|c|c|c|c|c|}
\hline $S_{red}$ & Entropy $S^{Potts}$ & Cross-entropy $S^{cross}$ & $\Delta S^{L_2}$ & $\Delta S^{AGG}$  \\
\hline  0.9& 62.97&60.01&-5.11& 8 .07\\
\hline  0.8& 75.18&62.88&-3.56& 15.86 \\
\hline  0.7& 81.07 & 58.65 &-2.03 &24.4\\
\hline  0.65& 82.96 & 56.27 &-1.68 &28.37\\
\hline  0.6& 83.81 & 51.43 &-0.67 &33.05\\
\hline  0.5& 86.58 & 45.48 &-0.31 &41.43\\
\hline
\end{tabular}
\end{center}
\caption{PF00014, release 27.0. Results of the Potts model inference with the ACE procedure. Potts states were grouped together according to their contributions to the site entropy, with cut-off $S_{red}$. Gaps are replaced with randomly drawn amino acids, see text.}
\label{table:spf00014Sred}
\end{table} 
 
\begin{table}
\begin{center}
\begin{tabular}{|c|c|c|c|c|}
\hline $p_{red}$ & Entropy $S^{IM}$ & Cross-entropy $S^{cross}$ & $\Delta S^{L_2}$ & $\Delta S^{AGG}$  \\
\hline 0.01& 52.38&50.61&-0.01& 1.78 \\
\hline  0.05 & 52.37 & 40.65&-0.01 & 11.73 \\
\hline
\end{tabular}
\caption{ PF00397, release 27.0. Entropies with  the independent Potts model and selection of Potts states based on their frequencies. Reweighting factor: $w=0.2$; regularization strength is $\gamma=0.002$.  }
\label{table:spf00397ind}
\end{center}
\end{table}

\begin{table}\begin{center}
\begin{tabular}{|c|c|c|c|c|}
\hline $p_{red}$& Entropy $S^{Potts}$ & Cross-entropy $S^{cross}$ & $\Delta S^{L_2}$ & $\Delta S^{AGG}$\\
\hline 0.01 & 38.25 &  39.76 &-3.25 & 1.78 \\
\hline 0.01, $\gamma =0.01$& 37.64 &  37.56 &-1.70 & 1.78 \\
\hline 0.05 &43.74& 32.67 &-0.66 & 11.73 \\
\hline 0.05, $\gamma =0.01$& 43.821 & 32.67 & -0.58& 11.73 \\
\hline
\end{tabular}
\caption{PF00397, release 27.0. Entropies with selection of Potts states based on their frequencies. Unless otherwise specified the regularization strength is $\gamma=0.002$. }
\label{table:spf00397pred}
\end{center}
\end{table}

\section{Application to phylogenetically related HIV proteins}\label{sec:hiv}

Human immunodeficiency virus (HIV) is distinguished by both a high mutation rate and a very short replication cycle, enabling the virus to rapidly generate mutations within specific parts of the viral sequence targeted by host immune responses, referred to as epitopes. Viruses bearing mutations within these epitopes are able to escape recognition by the immune system, thus preventing effective immune control of infection and allowing viral replication to continue unchecked. This process of rapid mutation also results in extraordinary HIV sequence diversity at the level of both individual hosts and of the virus population as a whole \citep{Korber:2001tc}. Together these factors contribute to the difficulty of designing an effective vaccine against HIV, which must be able to prime the immune system to combat diverse strains of the virus while also directing immune responses toward epitopes where escape is most difficult. Thus, quantifying the constraints on HIV proteins that limit their mutability represents an important step in the process of vaccine design.

Here, we estimate the entropy of various HIV proteins through maximum entropy models capturing the frequency of mutations at each site and the pairwise correlation between mutations. Previously, such models have been successfully used in the context of HIV to predict the fitness (ability to replicate) of a library of mutant strains of the virus \citep{Ferguson:2013kb,Mann:54ji}, and to explore aspects of protein structure and stability \citep{Haq:2012fw, Flynn:2015gv}. Unlike the protein families considered above, the HIV proteins we study here are more closely related phylogenetically, and thus the entropy that we compute may underestimate the true, potential variability of these proteins. However, this approach should still be successful in capturing nontrivial constraints on HIV. A variational mean-field theory calculation suggests that, while immune pressure due to genetically diverse individuals perturbs the inferred fields and phylogeny further modulates the fields in the maximum entropy model, the inferred couplings are not strongly affected \citep{Shekhar:2013je}.

The outline for this Section is as follows. In Section~\ref{sec:HIVproteome} we compare the entropy of all HIV proteins from different virus subtypes, except for the highly variable envelope subunit protein gp120. This analysis reveals a subset of strongly conserved HIV proteins that appear to be unusually constrained. In Section~\ref{sec:dS} we employ simulations to compute the loss in entropy when particular residues in these proteins are held fixed. There we show that the typical locations of escape mutations, which the virus uses to evade the host immune system, are associated with sites that strongly reduce the entropy when held fixed.

\subsection{Diversity across the HIV proteome} \label{sec:HIVproteome}
\begin{figure}
\begin{center}
\hspace*{-40pt}
\includegraphics[width=8.7cm]{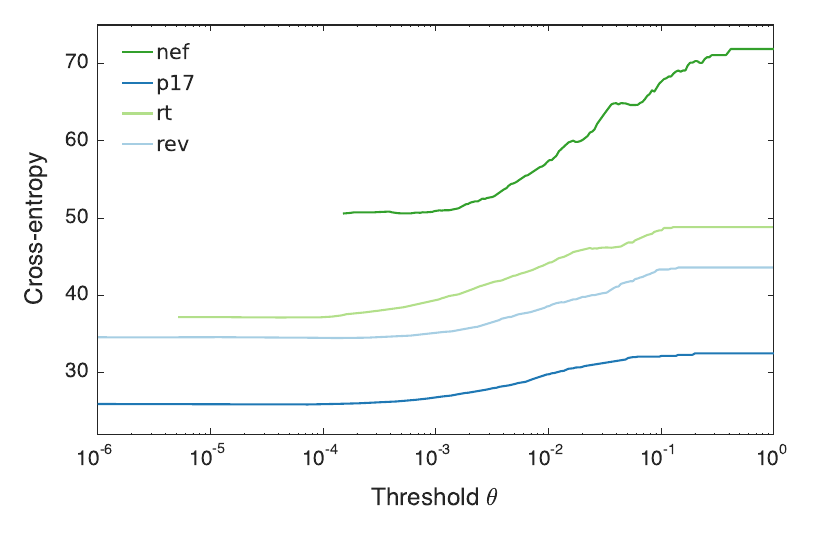}
\caption{{\small 
Typical behavior of the cross-entropy obtained by the ACE algorithm as a function of the threshold $\theta$, shown for various example HIV proteins. All exhibit good convergence toward stable values of the cross-entropy as the threshold is lowered. Similar results are also seen for the other proteins not shown here.
}
\label{fig:SvsT}}
\end{center}
\end{figure}

We inferred Potts models describing various HIV proteins from two prominent subtypes of the virus (clade B, dominant in Western Europe and the Americas, and clade C, dominant in Africa and parts of South Asia) using the adaptive cluster expansion method, and through this method we obtained an estimate of the entropy for each protein. For more details on the inference method and computation of the entropy, see Section~\ref{sec:ace}. The cross-entropy displayed good convergence for all proteins we tested, thus we do not expect large errors in our estimates of the entropy (Fig.~\ref{fig:SvsT}). In all cases we used an entropy cutoff of $S_{red}=0.9$, and regularization strength $\gamma\simeq 1/M$, where $M$ is the number of unique patients from which the sequence data was collected (ranges from approximately $500$ for some accessory proteins to $10,000$ for protease).

\begin{figure}
\begin{center}
\hspace*{-40pt}
\includegraphics[width=8.7cm]{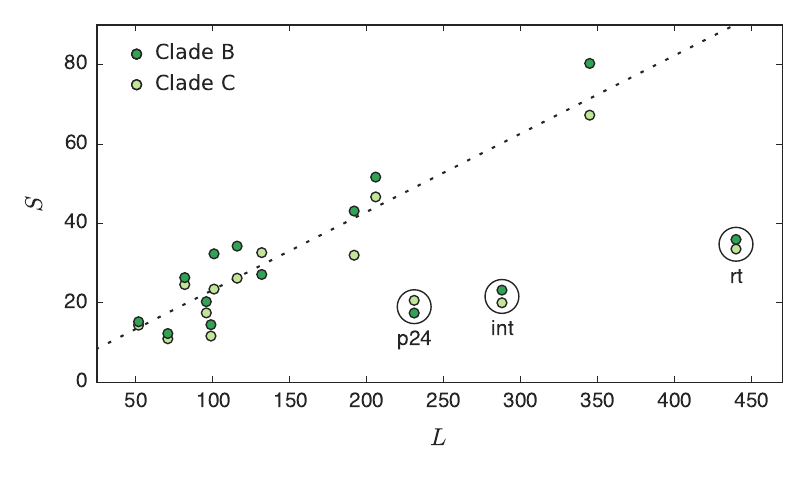}
\caption{{\small 
Entropy per site is comparable for most HIV proteins except for a subset that are more highly conserved. Here we show the entropy $S$ versus length $L$ for a set of HIV proteins. The typical entropy per site for these proteins, \emph{including} the effects of coupling between sites, is around $0.2$ (\textit{dotted line}). In contrast, the proteins p24, integrase (int), and reverse transcriptase (rt) appear to be substantially constrained, with entropy per site of only $0.08$ (\textit{circled}). Note that the surface protein gp120 is not included in this analysis; this protein is highly variable, and may exhibit higher entropy per site than typical HIV proteins.
}
\label{fig:SvsL}}
\end{center}
\end{figure}

In Fig.~\ref{fig:SvsL} we show the entropy $S$ of each protein versus its length $L$ in amino acids. We find that most HIV proteins have a typical entropy per site of around $0.2$, which holds for proteins obtained from both clade B and clade C viruses. Note that although the entropy scales roughly linearly with the protein length, this does not imply that variation at each site is independent; couplings between sites can contribute substantially to the entropy (see Fig.~\ref{fig:dS} below). In contrast with the typical scaling, there also exists a subset of HIV proteins that appear to be more highly constrained. Proteins p24, integrase, and reverse transcriptase have an entropy per site of roughly $0.08$, substantially lower than for other proteins.

There are several factors that may contribute to the reduced entropy observed for these proteins. At first, the reduced variability of p24 may appear surprising because this protein is frequently targeted by host immune responses \citep{Addo:2003ky,Streeck:2009et}, which would encourage frequent mutation. This protein forms the viral capsid, however, and is therefore subject to strict conformational constraints. The mature capsid is composed of p24 hexamers and pentamers that bind together in a ``fullerene cone'' shape \citep{zhao2013mature}. Previous work has shown that multiple mutations in residues along the p24 hexamer-hexamer interfaces, in particular, may be tightly constrained \citep{Dahirel:2011bi}. Epitopes in these regions are also frequently targeted by individuals who more effectively control HIV infection, possibly because mutations in these regions are more likely to damage viral fitness, thus decreasing the likelihood of escape \citep{Dahirel:2011bi}.

In contrast to p24, reverse transcriptase and integrase are not frequently targeted by host immune responses \citep{Addo:2003ky,Streeck:2009et}. They are responsible for the reverse transcription of viral RNA to DNA and the integration of viral DNA into the host genome, respectively. These proteins do not appear to be under substantial pressure to widely explore the sequence space \citep{Barton:2015uwa}, which, in addition to functional constraints, contributes to their reduced variability. Interestingly, we note that the conservation of reverse transcriptase observed here is also consistent with recent experimental studies that found extremely low tolerance for insertions in proteins involved in transcription for several different viruses \citep{Beitzel,Heaton10122013,Remenyi31102014,Fulton:2015uq}, suggesting that such proteins may potentially operate under strong functional constraints in more general cases.

\subsection{Relationship between the entropy and local pressure} \label{sec:dS}
In addition to characterizing the entropy for HIV proteins, we can also explore how variation at individual sites within a protein contributes to its overall entropy. The simplest way to do this is just to compute the single site entropy $S_{\rm site}(i)$ of each site $i$, obtained from the empirical correlations
\begin{equation} \label{eq:single-site}
S_{\rm site}(i) = -\sum_a p_i(a) \log p_i(a)\,.
\end{equation}
The drawback of this approach is that it neglects the effects of higher order constraints on protein sequences, such as those parameterized by the $J_{ij}(a,b)$, beyond just the frequency of amino acids observed at each site.

\begin{figure}
\begin{center}
\hspace*{-40pt}
\includegraphics[width=17.8cm]{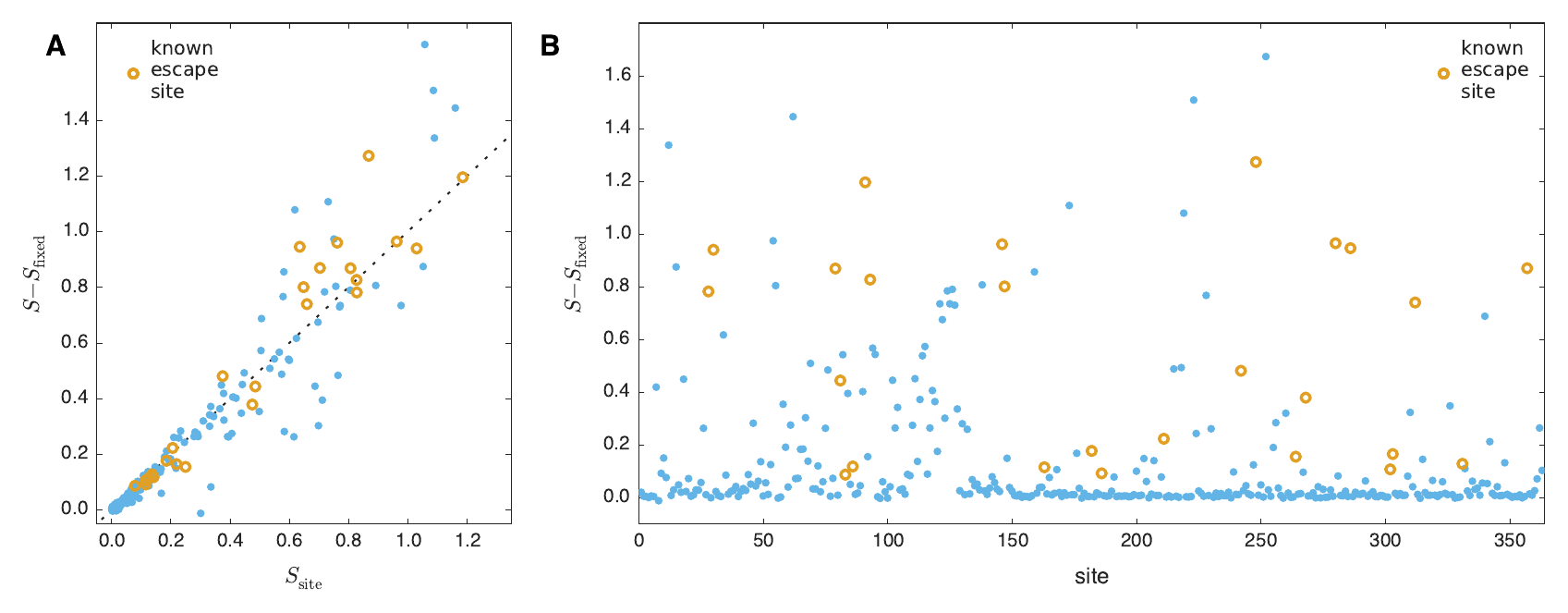}
\caption{{\small 
Change in entropy $S-S_{\rm fixed}$ upon individually fixing each site in HIV proteins p17 and p24 equal to the consensus amino acid. Known escape sites are highlighted (\textit{open circles}). ({\bf A}) Reduction in entropy from fixing a site is typically similar to the single site entropy $S_{\rm site}$, particularly for sites with low entropies, but sites with strong interactions can deviate significantly from this value. See main text for details. 
({\bf B}) Change in entropy as a function of position along the p17 (sites 1--132) and p24 (sites 133--363) proteins. Variation at known escape sites often contributes substantially more to the entropy than variation at other sites in the same epitope. Note that the CD8$^+$ T cell epitopes considered here are usually 9--11 amino acids in length. Escape mutations can occur at sites within the epitope or at nearby flanking sites.
}
\label{fig:dS}}
\end{center}
\end{figure}

To capture some of these higher order constraints, we can use the Potts model inferred for each protein to generate an ensemble of sequences with the amino acid at certain sites held fixed. We can then compute the entropy $S_{\rm fixed}$ of this ensemble of sequences using the adaptive cluster expansion method as before. The change in entropy $\delta S = S - S_{\rm fixed}$ upon fixing a site to a given value then quantifies the contribution of variation at that site to the entropy, including the effects of the inferred pairwise interactions. In the following, we choose to fix sites to their consensus values (one at a time), but the approach could be extended to any specific wild-type sequence. In Fig.~\ref{fig:dS}A, we see that the reduction in entropy from fixing most sites in the HIV proteins p17 and p24 is similar to the corresponding single site entropy $S_{\rm site}$. 
The effect of interactions is difficult to discern at this level for sites with very low variability. However, as shown in Fig.~\ref{fig:dS}A, $\delta S$ deviates substantially from $S_{\rm site}$ for a number of more variable sites where the effects of mutations are strongly coupled to other sites in the protein (note the scale in the above figure). 
The reduction in entropy for sites that lie above the line in Fig.~\ref{fig:dS}A is larger than expected from the single site entropy alone, indicating the presence of mostly positive (or, in the language of fitness, compensatory) couplings to other sites. For sites below the line $\delta S$ is smaller than expected, indicating more negative (or deleterious) couplings that tend to suppress mutation. These latter sites may then be good targets for effective immune responses.

Entropy has previously been associated with immune escape in HIV. Generally, escape tends to occur more rapidly at epitopes where the average single site entropy is high \citep{Ferrari:2011do,Liu:2013iu}. We also observe a connection between the sites where escape mutations are typically observed in well-characterized epitopes (see \citep{Ferguson:2013kb}) and entropy. In Fig.~\ref{fig:dS}B, we show the change in entropy upon fixing each site to the consensus amino acid in the p17 (sites 1--132) and p24 (sites 133--363) proteins, with known escape sites highlighted. Typically, these known escape sites contribute substantially more to the entropy than other sites within the same epitope. This result is intuitive: naturally we would expect that mutations at highly variable sites, or ones with many available compensatory interactions, should typically come at a low fitness cost to the virus, otherwise we would not frequently observe viruses with mutations at those sites. Mutations that both confer immune escape and which come at little fitness cost should then be selected more frequently.

\section{Exact and approximate values of the entropy for lattice-based proteins} \label{sec:LP}

\begin{figure}[htb]
\begin{center}
\includegraphics[width=10cm]{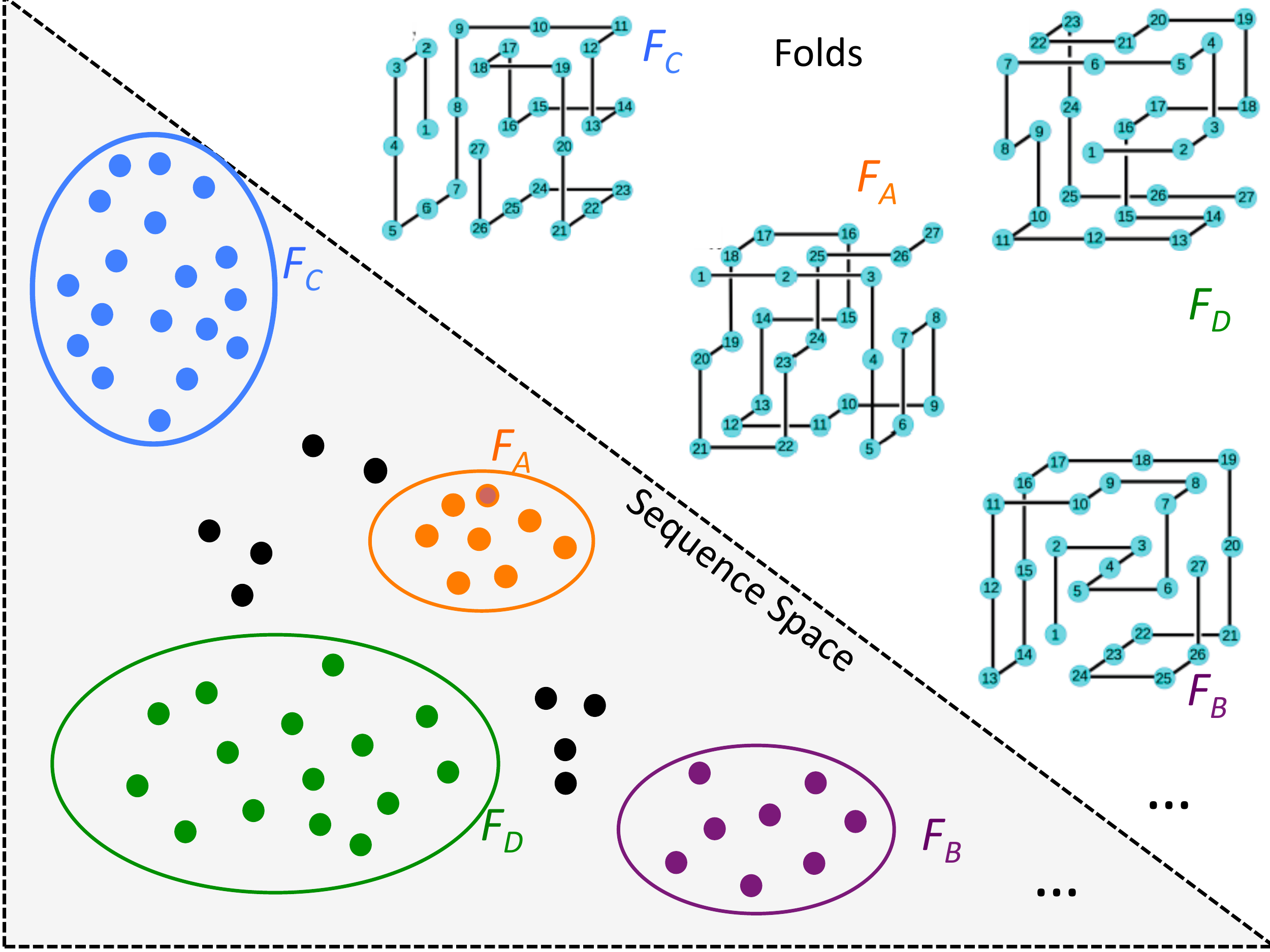}
\end{center}
\caption{ Pictorial representation of the sequence space (bottom left corner) and of four  depicted folds (top right corner)  among the $\mathcal{N}_{\rm fold}$ possible structures $F$. Sequences $\bA$ that fold in one of the four structures, say, $F$, {\em e.g.} such that $P_{\rm nat}(F|\bA)>0.995$, see (\ref{def_pnat}), are shown by coloured dots, with the same colors as the corresponding structures. Dark dots correspond to unfolded sequences, {\em i.e.} having low values of $P_{\rm nat}$ with all structures.  The logarithm of the volume in the sequence space associated to each fold defines its entropy, otherwise called designability \cite{li1996}. The entropies $S^{Potts}$ of the four folds shown in the figure  have been calculated in \cite{Jac:2015}, using the pairwise Potts models inferred from the families of sequences associated to the folds, with the ACE expansion and are recalled in Table~\ref{tab:MSA caracteristic}.  }
\label{fig:seqspa}
\end{figure}

\begin{figure}[htb]
\includegraphics[width=5cm]{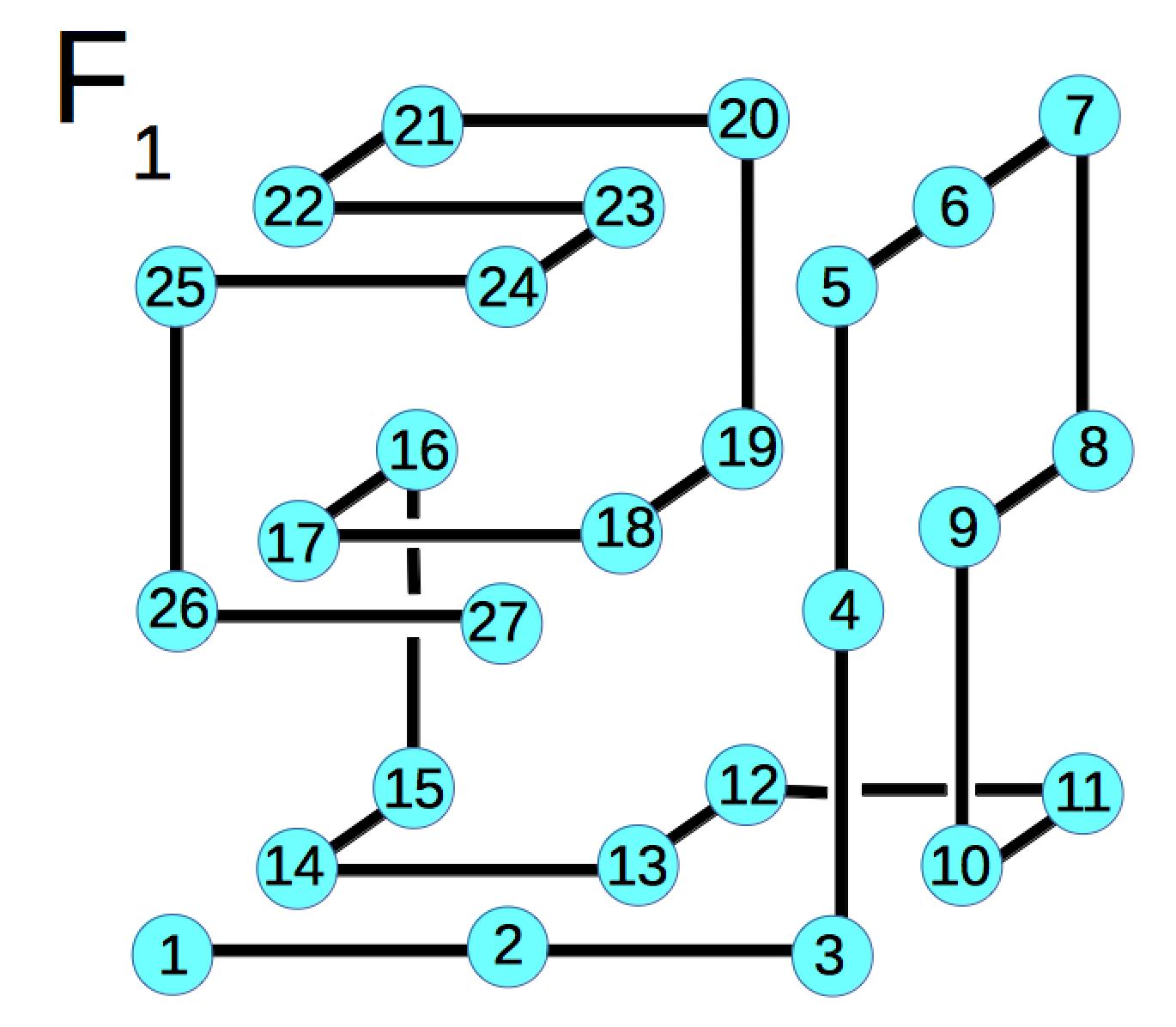}
\includegraphics[width=5cm]{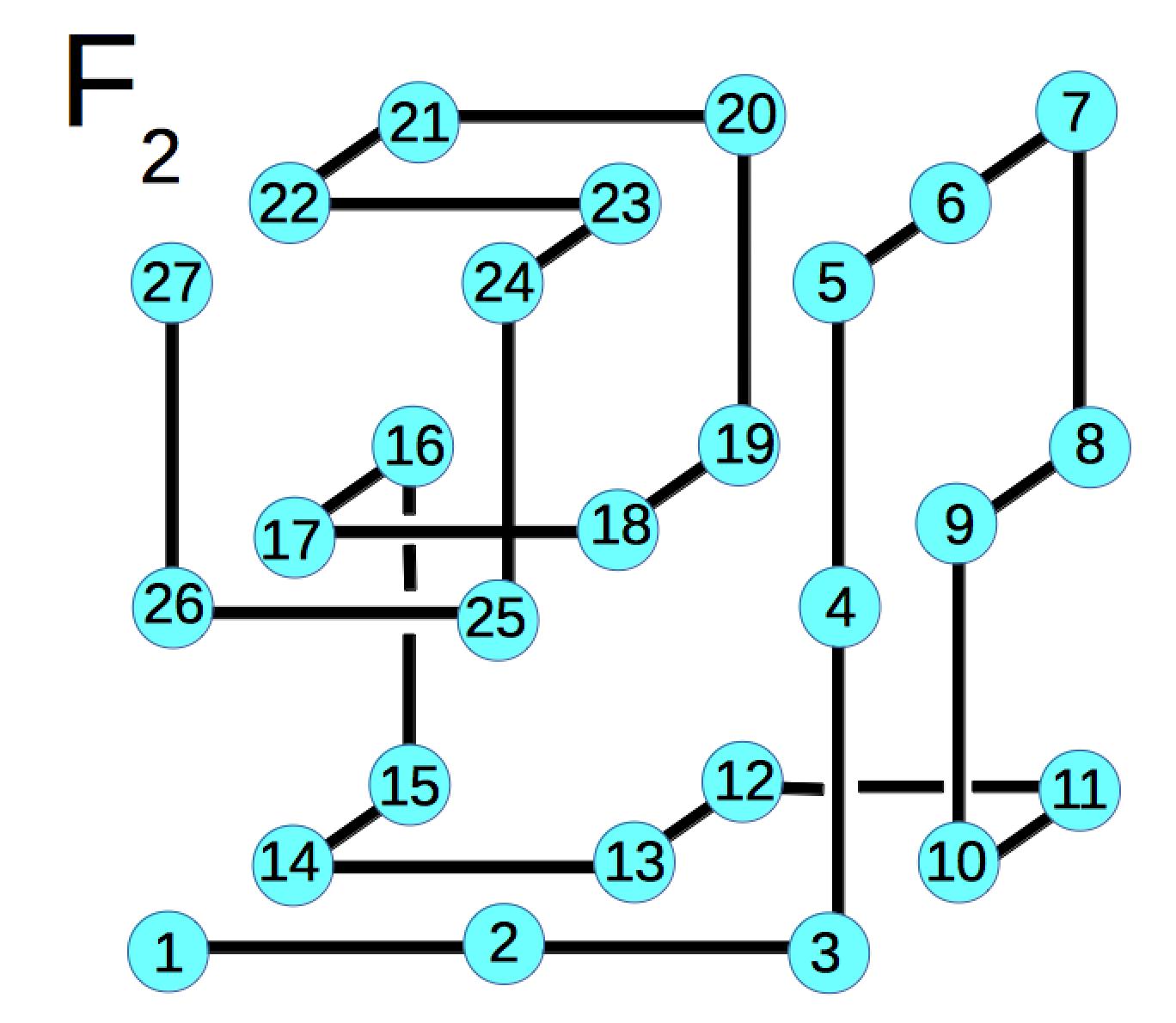}
\includegraphics[width=5cm]{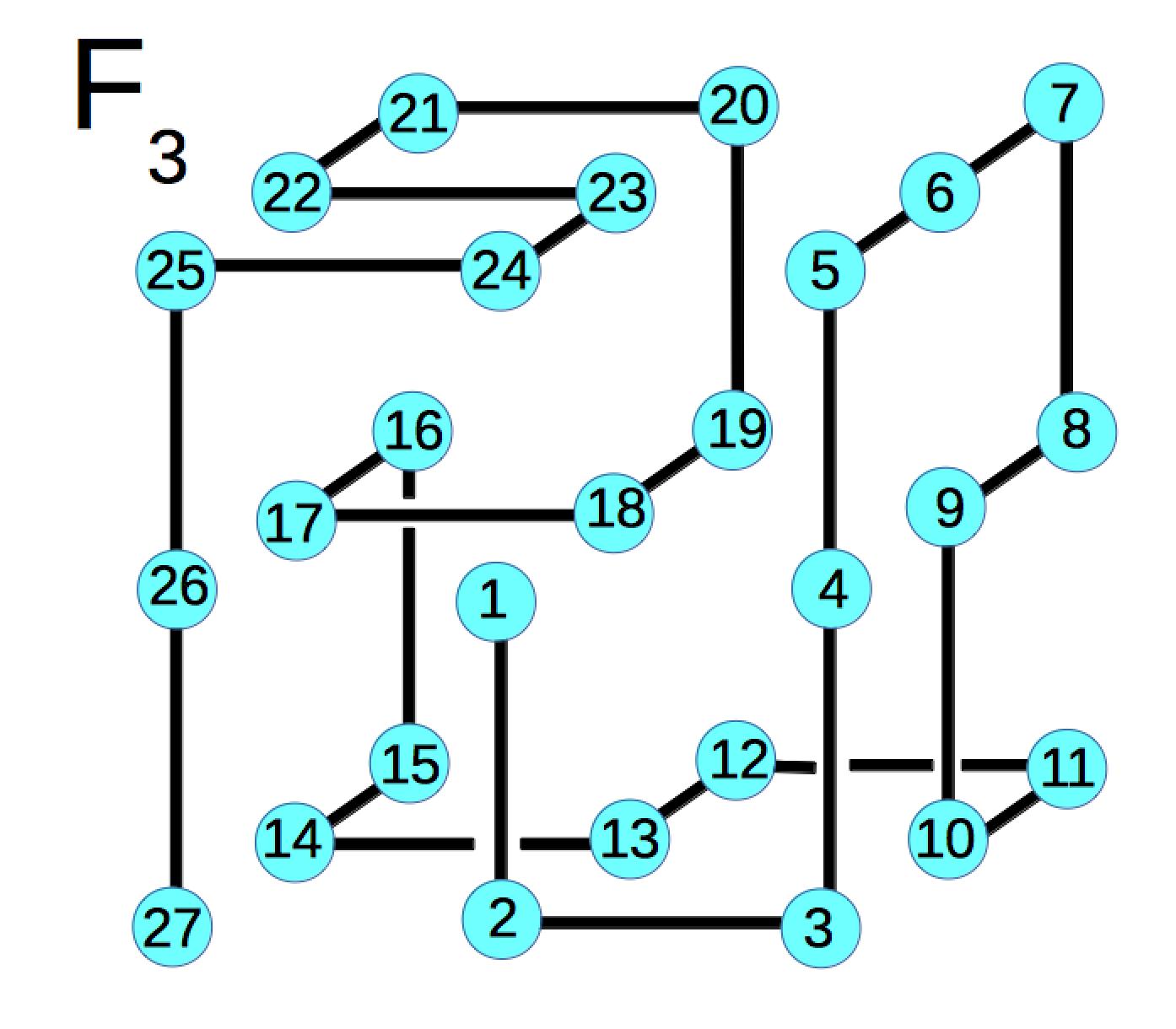}
\caption{The three folds considered here: $F_1$, $F_2$ and $F_3$. It is easy to see that $F_2$ is obtained from $F_1$ by simply exchanging the sites $25$ and $27$, while $F_3$ is obtained from $F_1$ by exchanging $1$ and $27$. We also see that the only sites affected by these exchanges are the nine sites $1,2,4,14,18,22,24,25$ and $27$.}
\label{fig:cubes}
\end{figure}

In this section, we compute the entropy of the families of lattice-based proteins (LP)  \cite{gutin90,li1996,li2002,england2003}.  Lattice proteins considered here  are composed of 27 amino acids occupying the sites of a $3\times 3 \times 3$ cube, see Figs.~\ref{fig:seqspa} and \ref{fig:cubes}. There are $\mathcal{N}_{\rm fold} = 103,346$ possible folds $F$ (conformations of the protein backbone) unrelated by symmetry. Given a fold $F$, each amino-acid sequence $\bA=(a_1,\ldots,a_{27})$ is assigned an energy
\begin{equation}
E_{LP}(\bA|F) = \sum_{i<j} c_{ij}^{(F)} E(a_i,a_j)
\end{equation}
where $E(a,b)$ is the Miyazawa-Jernigan statistical energy matrix \cite{Jernigan85}. The matrix $c^{(F)}_{ij}$ is the contact matrix
associated with the fold $F$: the entry is equal to unity if $i$ and $j$ are in contact, {\em i.e.}~are nearest neighbors on the cube, and zero otherwise.

The probability that a given sequence $\bA$ folds in conformation $F$ is defined following \cite{gutin90} as:
\begin{equation}
P_{\rm nat}(F|\bA) = \frac{\displaystyle e^{-E_{LP}(\bA|F)}}{\displaystyle \sum_{F'=1}^{\mathcal{N}_{\rm fold}} e^{-E_{LP}(\bA|F')}} = \frac{1}{1+\displaystyle \sum_{F' (\ne F)} e^{-\big[  E_{LP}(\bA|F')- E_{LP}(\bA|F)]\big]}} 
\label{def_pnat}
\end{equation}

From Eq.~(\ref{def_pnat}) it is clear that the  fold $F^*$,  which maximize the probability that a given sequence  $\bf A$ folds in it, is the one, among all  the other possible and competing structures $F'$, with minimal energy $E_{LP}({\bf A}|F^*)$.  However  the sequence is said to be folded in this structure $F^*$ only if $P_{\rm nat}(F^*|{\bf A})$  is very large, typically larger than 0.995. Therefore the requirement for  a sequence to fold  in a structure $F^*$ is  the existence of a large energy gap $E_{LP}(\bA|F')-E_{LP}(\bA|F^*)$ (at least of the order of five, in units of the temperature, set to unity here) with the other competing structures. Given a fold, this gap condition is generally satisfied by many sequences, see sketch in Fig.~\ref{fig:seqspa}, which define the protein family.  The fold attached to this set of sequences is called native fold, while the structures that have the smallest energy gap with the sequences in the set are said to be its close competitors.  

\subsection{Designability and entropy} 
An important characteristic of a structure is the volume (cardinality) of the set of attached sequences, see Fig.~\ref{fig:seqspa}, called designability \cite{li1996,england2003}.
The logarithm of the numbers of sequences folding in a given structure informally corresponds to the entropy defined here, see introduction. In \cite{li1996} it was shown by numerical simulations  that the designability depends on the structure: as sketched in Fig.~\ref{fig:seqspa}, some structures are associated to a large volume in the sequence space, while some correspond to smaller volumes. 
In \cite{england2003}, it was proposed that  the largest eigenvalue of the contact map $c_{ij}$ of a structure is indicative of its designability.

In a recent work \cite{Jac:2015},  large alignments of size $\mathcal{O}(10^4)$ for the four structures $(F_A, F_B, F_C, F_D)$ in Fig.~\ref{fig:seqspa}, were generated, and used  to infer the Maximum Entropy Potts models reproducing the 1- and 2-point statistics with the Adaptative Cluster Expansion described in Section \ref{sec:ace}. 
We summarize here the procedure we have followed to generate the alignments of sequences folding in the four structures of Fig.~\ref{fig:seqspa}, and the results we have obtained for their entropies. 
To generate a multi-sequence alignment  attached to a fold, say, $F$, we perform a search in the sequence space to find sequences $\bA$ with large folding probability $P_{\rm nat}(F|\bA)>0.995$  ~\cite{Ber:2007}. 
To this aim we have used a Monte Carlo procedure to sample the Gibbs distribution associated to the effective Hamiltonian
\begin{equation}
\mathcal{H}_W (\bA |F) = - \ln P_{\rm nat}(F|\bA) \ ,
\end{equation}
 in the sequence space at large inverse temperature ($\beta=10^{3}$). Here $W$ denotes the world of proteins, that is, the set of all possible structures; in \cite{Jac:2015} 10,000 folds among the $\mathcal{N}_{\rm fold}$  where randomly chosen. The sampled sequences form the multi-sequence alignment, which gives access to the 1- and 2-point statistics of the family.  We have then inferred the pairwise  Maximum-Entropy Potts model and computed its cross-entropy with the ACE procedure.
Results are given in Table~\ref{tab:MSA caracteristic}.

The Potts entropy is bounded from above by $27\times \log 20\simeq 80.9$; the difference between this upper bound (corresponding to a set of  $L=27$ fully unconstrained amino acids) and the Potts entropy is a measure of the structural constraints acting on the sequences.
 As reported in Table~\ref{tab:MSA caracteristic}  we have also compared the Potts entropy to different estimators,
such as the maximal eigenvalue of the contact matrix ${\bf c}^{(F)}$ of the target fold under consideration \cite{england2003}, and the mean sequence variability in the alignment (average Hamming distance to the consensus sequence across the alignment), see Supplementary Information in \cite{Jac:2015}.  The general picture that arises from \cite{Jac:2015} is that the presence of competing folds that are close (either in terms of the contact matrix or in terms of energy gaps) to the native fold globally constrains the protein sequences and reduces the entropy of the family, hence defining an entropy cost associated to the competition in folding. Hereafter we show that this cost can be accurately computed in the case of a very small number of competing structures. This simple `protein world' can be used, in turn, as a testbed for the inference algorithm and the approach developed in Section~\ref{sec:ace}.
 
\begin{table}[h]
\begin{center}
\begin{tabular}{|c|c |c|c|c|}
\hline {\bf Fold} & {\bf Top eigenvalue}   & {\bf Potts Entropy}  & {\bf Mean \%}& {\bf Dist. to}  \\
  & {\bf of} {$\bf c$}  &   {\bf (ACE)} & {\bf identity btw seq.}&{\bf nearest struc.  } \\
\hline $F_B$ & 2.6 & 50.2 & 24 & 14\\
\hline $F_A$ & 2.5 &  50.9 & 23 & 11\\
\hline $F_D$ & 2.7  & 55.4 & 21& 9\\
\hline $F_C$  & 2.9  & 58.4 & 19& 4\\
\hline
\end{tabular}
\end{center}
\caption{Estimates of how designable are the proteins families associated to structures $F_A,F_B,F_C,F_D$ (ranked in increasing order of their entropies):  largest eigenvalues of the contact map matrix $\bf c$ and of the corrected matrix ${\bf c-\bar c}$ (1st column), entropy of the inferred Potts model obtained by  ACE (2nd), and mean percentage of identity between sequences (3rd). We also give the distance to the nearest structure (4th column).  For the identity calculation, we  average the number of amino acids that take their consensus  values, and divide by the 
number of amino acids in the protein ($=27$). }
\label{tab:MSA caracteristic}
\end{table}%

\subsection{Exact calculation of the entropy for pairs or triplets of proteins}\label{sec:lpexact}

We start from the simplest case, that of a unique possible fold, $F_1$ in Fig.~\ref{fig:cubes}. In that case, any sequence $\bA$ will necessarily fold into $F_1$, and the corresponding effective Hamiltonian ${\cal H}_{F_1} (\bA|F)$ vanishes. The amino acids can be assigned randomly on each site, and the entropy is simply (Table ~\ref{tab:entro}, top line):
\begin{equation}\label{sf1}
S (F_1) = \ln \big(20^{27}\big) = 80.8848 \, .
\end{equation}
In a more complex protein world made of two proteins, $F_1$ and $F_2$, the probability that a sequence $\bA$ folds into $F_1$ now defines the effective Hamiltonian:
\begin{equation}\label{p0}
\mathcal{H}_{ [F_1 ; F_2]}(\bA|F_1) = - \log \left( 1 + e^{-E_{LP}(\bA|F_2)+E_{LP}(\bA|F_1)} \right)
\label{H12}
\end{equation}
where $[F_1 ; F_2]$ denotes the two-protein world made of $F_1$ and $F_2$, with $F_1$ chosen as the reference fold.
On our small cube, the contact matrices are uniquely defined by a set of $28$ contacts (pairs of neighbours on the cube, excluding contiguous sites on the protein backbone). We have found a large number of pairs of protein folds that share $24$ out of $28$ of those contacts.  Choosing $F_1$ and $F_2$ to have $24$ common contacts (Fig.~\ref{fig:cubes}), we have only $4$ pairs of sites that are relevant in the calculation of the energy difference in (\ref{p0}). The effective Hamiltonian will be constraining $8$ sites (2 for each contact) at most, and will not depend on the amino acids on the other sites. It turns out that out of those $8$ sites, the 4 differing contacts are carried by only 6 distinct sites. The calculation of the partition function associated to $\mathcal{H}_{ [F_1;F_2]}(\bA|F_1) = \mathcal{H}_{[F_1;F_2]}(a_1,a_2,...,a_6|F_1)$ is numerically tractable as it involves a summation over $20^6$ configurations only,
\begin{equation}
 \displaystyle Z _{ [F_1;F_2]}= 20^{21} \sum_{a_1=1}^{20} \cdots \sum_{a_6=1}^{20} ~ e^{- \beta\mathcal{H}_{ [F_1;F_2]}(a_1,\ldots,a_6|F_1)} 
 \ ,
 \end{equation}
 and the corresponding entropy for the fold $F_1$  is
\begin{equation}
 S \big( [ F_1;F_2 ])= \ln Z _{ [F_1;F_2]}-  \frac{d }{d\beta} \ln Z _{[F_1;F_2]} \ .
\end{equation}
The value of this entropy is given in Table \ref{tab:entro} (1st column), and is close to $77.16$. The decrease with respect to $S(F_1)$ in (\ref{sf1}) measures the loss in entropy due to the introduction of the competing fold $F_2$. Using a conversion in log base $20$ the entropic cost is of $\approx 1.3$ site. 

We then consider another fold $F_3$. This third structure is also close to $F_1$, see Fig.~\ref{fig:cubes}, and a bit further away from $F_2$.
We have calculated the entropy of the two-fold world comprised of $F_1$ and $F_3$: we find that $S \big( [F_1 ; F_3] \big)$ is identical to $S \big( [ F_1;F_2 ])$ as $F_2$ and $F_3$ both share 24 contacts with $F_1$. The entropy $S \big( [F_2 ; F_3] \big)$ is slightly larger than $S \big( [F_1 ; F_3] \big)$ (Table \ref{tab:entro}), as can be expected from the fact that $F_2$ and $F_3$ are further apart, with only 22 common contacts. The energy gap between $F_2$ and $F_3$ is therefore larger than between $F_1$ and $F_3$, and sequences folding in $F_2$ are less constrained by the presence of the competing fold $F_3$ than the sequences folding in $F_1$ in the presence of the competing fold $F_3$ too. This result agrees with the qualitative findings of \cite{Jac:2015}.

When this third fold $F_3$ is added to the protein world,  
the effective Hamiltonian (associated to the folding into $F_1$) reads
\begin{equation}
\mathcal{H}_{[F_1;F_2 F_3]} (\bA|F_1) = - \log \left( 1 + e^{-E_{LP}(\bA|F_2)+E_{LP}(\bA|F_1)} + e^{-E_{LP}(\bA|F_3)+E_{LP}(\bA|F_1)} \right)
\label{H123}
\end{equation}
and depends on the values of nine amino acids on the sequence only. The calculation of the partition function and of the entropy $ S \big(  [F_1;F_2 F_3] \big)$ can be done along the lines above; it now requires to sum up over $20^9$ configurations, and was done in one CPU day on a desktop computer.  Addition of a third fold leads to a value of the entropy of $75.37$, which shows that an additional $0.8$ site has been constrained in the process, see Table~\ref{tab:entro}.

\begin{table}[htb]
\begin{center}
\begin{tabular}{|c|c|c|c|c|}
\hline
{\bf Protein world} & {\bf Exact} & {\bf Ind. Model} & {\bf Potts (ACE)} & {\bf Potts (Exact)}  \\
\hline
 $F_1$ & 80.8848 & 80.8848 & 80.8848 & 80.8848 \\
\hline
 $[F_1 ; F_2]$ & 77.1560  & 77.5035 &  77.1060 & 77.2504 \\
\hline
$[F_1 ; F_3]$ & 77.1560  & 77.5035 &  77.1060 & 77.2485 \\
\hline
$[F_2 ; F_3]$ & 77.2054 &  77.8174 &  77.2294 &  \\
\hline
$[F_1; F_2 F_3]$& 75.3762 &  75.7432 &  75.3331 &  \\
\hline
\end{tabular}
\end{center}
\caption{Entropies for the family associated to the fold $F_1$ and for the protein worlds with one, two and three structures, as calculated exactly, by fitting an independent model or by fitting a Potts model, either with the ACE algorithm or with exact calculation. For the protein world $[F_2;F_3]$, the entropy is that of the family associated to the fold $F_2$. Empty cells signal entropies that would have been to costly to compute, see main text. }
\label{tab:entro}
\end{table}

\subsection{Approximate calculation of the entropy based on the inferred Potts models}

As the models above are exactly solvable, they can be used as a simple testbed for our Maximum Entropy-ACE approach, we have already used for real protein data. To do so, we 
have computed the one- and two-point marginals, $p_{i}(a)$ and $p_{ij}(a,b)$ for the protein worlds $[F_1;F_2]$, $[F_1 ; F_3]$, $[F_2 ; F_3]$ and $[F_1;F_2 F_3]$, from very large MSA with $5\times 10^5$ sequences generated through Monte Carlo sampling. We first fit the 1-point statistics only with an Independent Model (IM). The corresponding
entropies are given by (\ref{defsim45}), with $q_i=20$ for all 27 sites, with values listed in the second column of Table~\ref{tab:entro}. 

\begin{figure}[htb]
\begin{center}
\includegraphics[width=8cm]{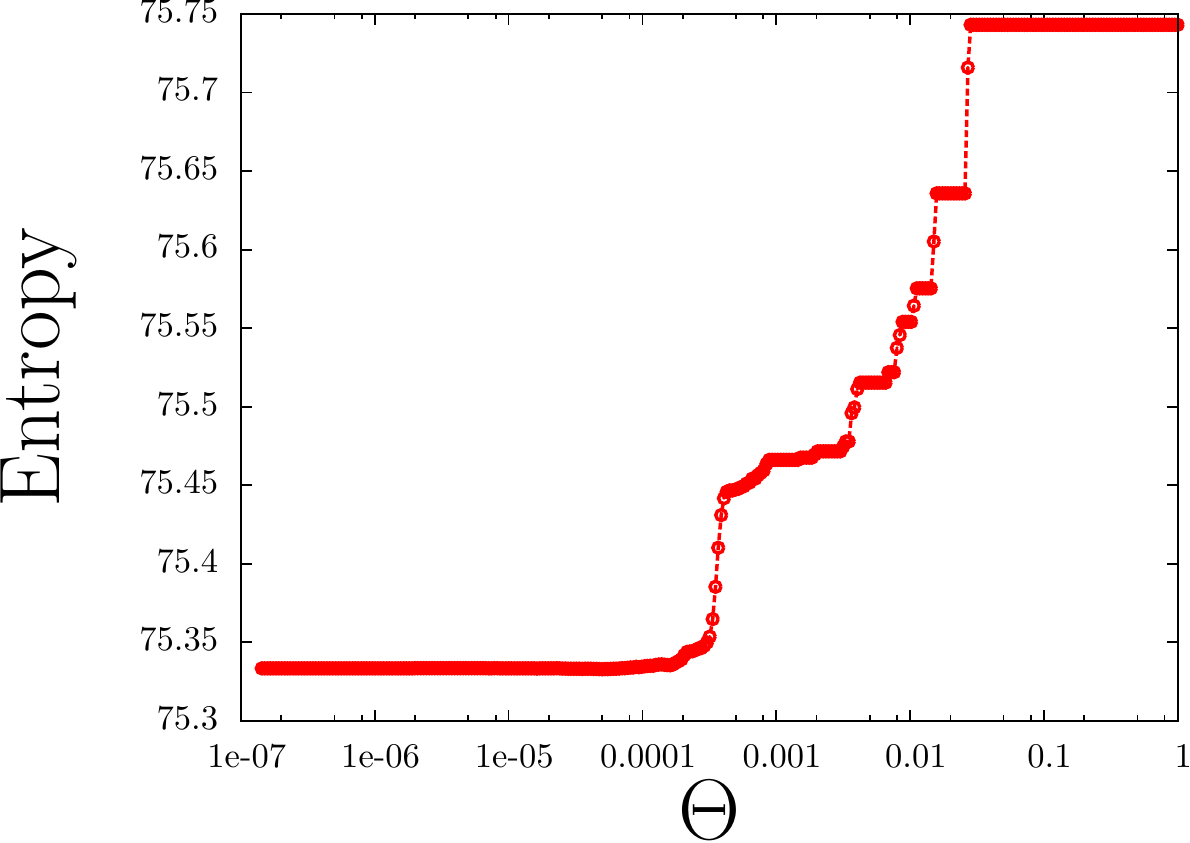}
\end{center}
\caption{Entropy of the protein family associated to $F_1$ as a function of  the threshold $\theta$ as computed by the ACE procedure in the case $[F_1; F_2  F_3]$. The entropy saturates to a value very close to the exact one, see Table~\ref{tab:entro}. The plateau at the beginning of the calculation (large $\theta$) corresponds to the entropy of the independent model (IM).}
\label{fig:cluster_entro2}
\end{figure}

We  then take into account the 2-point statistics, and infer the corresponding Maximum-Entropy Potts models with the ACE algorithm.  We show in Fig.~\ref{fig:cluster_entro2} the behaviour of the entropy predicted by the ACE algorithm for the case $[F_1; F_2 F_3]$, as a function of the threshold $\theta$ in the algorithm (see Section~\ref{sec:ace}). Similar curves are obtained for the protein worlds made of two structures. The entropy converges to a value very close to the exact value calculated through complete enumeration of the Potts states, see Section~\ref{sec:lpexact}. Even though our sampled alignment is very large here, correlations are still not exactly measured, leading to seemingly significant correlations on pairs of sites outside the restricted subset involved in the exact partition function. Due to those spurious constraints, the entropy is slightly lower than its exact value (Table~\ref{tab:entro}, third column).

Last of all, we can determine the coupling parameters of the Potts model by brute force optimization of the cross-entropy \ref{eq:Pottsentropy}, as the number of sites effectively involved is small (see discussion above).  This computation provides us with the exact entropy of the Potts model associated to the MSA we have generated, see results in Table~\ref{tab:entro}, fourth column. The computation  takes about one hour on a desktop computer when the number of relevant sites is $6$ (for the worlds $[F_1 ; F_2]$ and $[F_1 ; F_3]$), but would require several days when the number of relevant sites is $9$ (for the worlds $[F_2 ; F_3]$ and $[F_1 ; F_2 F_3]$). As expected, the entropy of the exact Potts model is now larger than the exact entropy (Table~\ref{tab:entro}, first column): the Potts model is, indeed, less constrained than the many-body model defined by the Hamiltonians in (\ref{H12},\ref{H123}).

\section{Discussion}\label{sec:discussion}

In this paper we have used different methods to calculate the entropy of protein families. One of our main findings, obtained from the wide-scale comparative analysis of a large number of PFAM families, is that the entropy primarily depends on the length $N$ of the family profile. More precisely, we find a linear scaling  $S\simeq \sigma N$. The value of the slope $\sigma$ depends on the method we have used. For the HMM model we find $\sigma\simeq 1.9-2.2$. Maximum Entropy modelling of a few protein families with pairwise interaction Potts models give values of  $\sigma$ ranging between 1.2  (when all amino acids present in the multiple sequence alignment are kept in the modelling) to 1.7 (for large reduction in the number of a.a. used), while the independent-site model give $\sigma \simeq 1.7-1.8$. 

Those estimates for $\sigma$ are compatible with previous results in the literature. The authors of \cite{Sha:1998} estimated $\sigma \simeq 1.9$ based on the following modelling of a protein family. Given the contact map ${\bf c} =\{c_{ij}\}$ of the native fold corresponding to the family (supposed to be perfectly conserved throughout the family), the energy of an amino-acid sequence $\bA$ is approximated as a sum of pairwise energetic parameters between amino-acids in contact on the structure (relative distance smaller than $6.5$~\AA),
\begin{equation}
 E_{AP}(\bA, {\bf c})=\sum_{i<j} \, E(a_i,a_j) \; c_{ij} \;
\end{equation}
The energetic parameters $E(a,a')$ describe the magnitude of the interaction between amino acids $a$ and $a'$, and are given by the Miyazawa-Jernigan energetic matrix; variants of this  statistically derived energy matrix $E$ were proposed without affecting much the value of $\sigma$. The Gibbs distribution associated to this energy is the sequence distribution for the family. By computing the average energy $\langle E\rangle (T)$ at different temperatures $T$ with Monte Carlo simulations, one can obtain the value of the entropy through thermodynamic integration:
\begin{equation}
S(T)-S(\infty)=\frac{\langle E\rangle (T)}{T}-\int_{T}^{\infty} dt\, \frac{\langle E\rangle (t)}{t^2} \ .
\end{equation}
In the formula above, $S(\infty)$ is the entropy of the system at infinite temperature, and is equal to $N$ times the entropy of the background amino acid distribution, $s_{BG}=-\sum _{a=1}^{20} p(a) \log p(a)$. As a result the estimate of $\sigma \approx 1.9$ was found, see Fig.~2 of \cite{Sha:1998}.

The entropies we have found with the HMM models are larger than with the Maximum Entropy Potts approach. One possible explanation is that, while HMM are routinely used to identify families, they are not supposed to reproduce faithfully the statistics of the multi-sequence alignments (MSA) when used as generative models. More precisely, HMM generate sequences that are more variable than the ones found in natural MSA, even at the level of single-site frequencies. In the Maximum Entropy Potts approach, we find smaller values of the entropy, especially when increasing the number of Potts states on each site (up to the number of amino acids observed at least once in the MSA). The reason is that increasing the number of pairwise correlations to reproduce corresponds to increasing the number of constraints to satisfy, and therefore leads to a decrease in entropy. However, this may also lead to overfitting the data if the number of sequences in the MSA is too small. 
  
In the case of phylogenetically related HIV sequences we find  a ten-fold decrease for the entropy per site, $\sigma\simeq 0.2$.  This small value reflects the high phylogenetic correlations between sequences and the poor variability in the MSA. To better understand how this value compares to the ones we have found for protein families, we have considered the  example of the RT (reverse-transcriptase, PF00078), a long protein with more than 500 amino-acids, which is unusually conserved in the HIV data (entropy per site $=0.08$). We have looked at one domain of this RT protein, known as PF06817, the so-called RT thumb domain, composed of a four-helix bundle. In HIV data,  the first 10-15 sites of the domain, not counting gaps, tend to be quite conserved, while the latter part is more variable. The resulting entropy is very low. Conversely, the HMM profile shows much less conservation. The full alignment on the PFAM database contains sequences from many different viruses, so this might also contribute to the observed variability (especially if viewing the representative proteomes on PFAM). It might be the case that, while (at least part of) this protein is well conserved in HIV, it is not as conserved across many different viruses. RT  is not thought to be often targeted by human immune responses, so that will contribute to the reduced variability in HIV, in addition to functional constraints. Intuitively we would expect that this protein as a whole should be functionally constrained, but perhaps either the constraints are virus-specific, or the frequency of targeting by the immune factor is the dominant reason why it appears more conserved.

While the entropy computed in the presence of high phylogenetic correlations is not representative of the diversity in the protein family which may be observed across distant organisms, it can be used to characterize the constraints acting  on the different sites of a given protein, on the different proteins of the HIV virus. In particular we have computed the cost in entropy corresponding to fixing the amino acid content on one site, {\em e.g.}~to its consensus value. While this cost is close to the entropy of the single-site amino-acid frequencies for most sites, the two quantities differ on some sites, which signals the presence of strong coupling effects (epistasis). Computing the entropy cost offers another potential avenue to investigate the fitness landscape of the virus. Sites associated to high entropies are likely to be the sites of escape mutations for the virus, in response to host immune pressure. Note that, from a computational point of view, the limited variability in the MSA helps for the inference of the Potts model. The system is, in physical jargon, in a paramagnetic phase with large local fields, and the Independent Site Model already provides a good starting point for the inference.
  
In the artificial lattice-based protein models we have studied, the entropy is very large, $\sigma\simeq 3$, due to the extremely reduced protein worlds we have considered (only a few proteins coexist and compete), in order to be amenable to exact calculations. Calculations taking into all the possible competing structures on the cubic lattice show a drastic reduction in the entropy per site, and give  $\sigma\simeq 1.8-2.1$ \cite{Jac:2015}, a value close to the one found for real protein families. It is important to underline that, while the lattice-protein model does not contain only 2-body interactions, the true entropy is very accurately recovered with the pairwise Potts model, see Table~\ref{tab:entro}.   
 
An important question is whether our values for the entropy can be confronted to experiments. In directed evolution experiments, starting from a pool of random sequences, sequences are selected according to their {\em  in vitro} fitness, such as binding affinity against a target. The fittest sequences are mutated, amplified, and another round of selection can take place. One fundamental issue is the size of the initial pool of sequences allowing for the selection of (at least one) fit protein(s). In one experiment \citep{keefe:2001} Keefe and Szostak started from a pool of $6\times 10^{12}$ proteins with 80 amino acids each, and selected them according to their ATP binding affinity. After 4 cycles of selection and mutation (made possible by the RNA tags attached to the proteins) they found 4 different sequences of new ATP binding proteins. The authors estimate that 1 in $10^{11}$ random sequences has ATP-binding activity comparable to the one isolated in the study. Assuming that this ratio corresponds to the ratio of the number of proteins in the `ATP-binding family' over the number of sequences with 80 amino acids, we obtain that the entropy of this putative family is $S = \ln (10^{-11} \times 20^{80}) \simeq 214.3$. The entropy per site is therefore $\sigma \simeq 2.67$. This estimate is large compared to the values we have found in the analysis of the natural protein families in the present work. One possible explanation is that the definition of `ATP-binding family' is actually too loose compared to natural families, which would lead to high apparent values for the entropy. We believe that further work to connect estimates of the entropy and {\em in vitro} directed evolution experiments in a quantitative way would be very useful.
 
Last of all, while we have considered here the entropy of the distribution of amino acid sequences, we should not forget that those sequences are coded at the DNA level by nucleotides. The redundancy of the genetic code adds extra entropy to the value we have computed. This additive contribution depends on the amino acid content, as the degeneracy of amino acids varies from 1 to 6. In addition, it also depends on the organisms and on the tissue where the protein are expressed through the codon bias. More subtle effects, {\em e.g.}~resulting from the pressure exerted by the innate immune system, also limit the diversity of the nucleotide sequences at fixed amino-acid content \cite{greenbaum2014}.

  \vskip .3cm \noindent
  {\bf Acknowledgements.} S.C., H.J. and R.M. were partly funded by the Agence Nationale de la Recherche Coevstat project (ANR-13-BS04-0012-01).


\bibliographystyle{unsrt}

\bibliography{entro}

\end{document}